\pdfoutput=1 
\documentclass[onecolumn,12pt]{article}
\usepackage{jheppub} 
\usepackage{xfrac}
\usepackage{subfigure}
\usepackage{amsfonts}
\usepackage{verbatim}
\usepackage{amssymb}
\usepackage{graphicx}

\usepackage[utf8]{inputenc}
\usepackage[T1]{fontenc}
\usepackage{overpic,color}

\usepackage{epstopdf}
\usepackage{mathrsfs}
\usepackage{amsfonts}
\usepackage{amsmath}
\allowdisplaybreaks[4]        
\usepackage{amssymb}
\usepackage{euscript}     
\usepackage{color}         
\usepackage{tensor}        
\usepackage{amsthm}
\usepackage{graphicx}
\usepackage{subfigure}
\usepackage{bbm}
\usepackage[header,title,page,titletoc]{appendix}  
\usepackage{float}

\usepackage[numbers]{natbib}  
\usepackage{float}

\usepackage{color}
\usepackage[dvipsnames]{xcolor}

\definecolor{darkblue}{rgb}{0.0, 0.0, 0.55}
\definecolor{grey}{rgb}{0.57, 0.64, 0.69}
\definecolor{lightbrown}{rgb}{0.71, 0.4, 0.11}

\renewcommand{\(}{\left(}
\renewcommand{\)}{\right)}
\renewcommand{\[}{\left[}
\renewcommand{\]}{\right]}

\newcommand{\be}{\begin{equation}}
\newcommand{\ee}{\end{equation}}
\newcommand{\bea}{\begin{eqnarray}}
\newcommand{\eea}{\end{eqnarray}}
\newcommand{\beq}{\begin{equation}}
\newcommand{\eeq}{\end{equation}}
\newcommand{\beqa}{\begin{eqnarray}}
\newcommand{\eeqa}{\end{eqnarray}}
\newcommand{\beqar}{\begin{eqnarray*}}
\newcommand{\eeqar}{\end{eqnarray*}}

\newcommand{\pa}{\partial}

\title{\centering\boldmath Hairy Black Hole Chemistry}

\author[a]{Dumitru Astefanesei,}
\author[b,c]{Robert B. Mann,}
\author[a]{and Ra\'ul Rojas}

\affiliation[a]{Pontificia Universidad Cat\'olica de Valpara\'iso, Instituto de F\'isica, \\ Av. Brasil 2950, Valpara\'iso, Chile.}
\affiliation[b]{Department of Physics and Astronomy, University of Waterloo, \\ Waterloo, Ontario, N2L 3G1, Canada}
\affiliation[c]{Perimeter Institute, 31 Caroline Street North, \\ Waterloo, ON, N2L 2Y5, Canada}

\emailAdd{dumitru.astefanesei@pucv.cl}
\emailAdd{rbmann@uwaterloo.ca}
\emailAdd{raulox.2012@gmail.com}

\abstract{
We study the thermodynamics of an exact hairy black hole solution in Anti-deSitter (AdS) spacetime. We use the counterterm method supplemented with boundary terms for the scalar field to obtain the thermodynamic quantities and  stress tensor of the dual field theory. We then extend our analysis by considering a dynamical cosmological constant and verify the isoperimetric inequality. Unlike the thermodynamics of Reissner-N\"{o}rdstrom (RN) black hole in  this `extended' framework, the presence of the scalar field and its self-interaction make also the criticality possible in the grand canonical ensemble. In the canonical ensemble, we prove that, in fact, there exist two critical points. Finally we comment on a different possible interpretation that is more natural in the context of string theory.}

\begin{document} 

\maketitle
\flushbottom

\section{Introduction} 
\label{intro}

Hairy black hole solutions in gravity models with a negative cosmological constant have attracted considerable interest due to their essential role in understanding various properties of field theories with a holographic dual \cite{Maldacena:1997re,Aharony:1999ti} (see, also, \cite{Nastase:2017cxp,Ammon:2015} and references therein). The scalars can also cluster to form smooth horizonless compact objects in AdS, the so-called `boson stars' \cite{Astefanesei:2003qy,Astefanesei:2003rw,Buchel:2013uba}
and can yield black hole spacetimes with a single Killing vector exhibit an interesting range of behaviour
\cite{Dias:2011at,Stotyn:2011ns,Stotyn:2012ap,Gonzalez:2013aca,Henderson:2014dwa,Brihaye:2014nba,Herdeiro:2015kha}.

Over the past decade, the framework of black hole thermodynamics was extended to allow  for a `dynamical pressure' and its conjugate volume.
That gave rise to the `black hole chemistry' program \cite{Kubiznak:2014zwa} where the cosmological constant in Anti-deSitter (AdS) spacetime
taken to be a thermodynamic variable \cite{Henneaux:1985tv, Creighton:1995au} that is interpreted as a pressure term \cite{Kastor:2010gq, Kastor:2011qp}.  It turns out that this volume seems to possess some universal properties and is conjectured to satisfy a relation called the Reverse Isoperimetric Inequality
\cite{Cvetic:2010jb}  whose violation has recently been conjectured to correspond to a new kind of
instability for black holes  \cite{Johnson:2019mdp}.

There is now a large body of research that has been carried out  in this area
 \cite{Kubiznak:2016qmn}, whose key finding is that  black holes can display a broad range of phase behaviour that has counterparts in other areas of physics. This began with the understanding that 
there is a deep analogy between charged anti-de Sitter black holes and  Van der Waals fluids \cite{Kubiznak:2012wp,Gunasekaran:2012dq} beyond that originally posited in the context of the AdS/CFT correspondence
\cite{Chamblin:1999tk,Chamblin:1999hg}.  Further examples were then found 
including re-entrant phase transitions \cite{Altamirano:2013ane,Frassino:2014pha}, triple points \cite{Altamirano:2013uqa,Frassino:2014pha}, polymer-like behaviour \cite{Dolan:2014vba}, 
and even superfluid-like phase transitions \cite{Hennigar:2016xwd,Hennigar:2016ekz,Dykaar:2017mba}.
Black holes can also be understood as heat engines \cite{Johnson:2014yja,Chakraborty:2016ssb,Hennigar:2017apu,Chakraborty:2017weq,Johnson:2019olt,Ahmed:2019yci},  and have a sensible Joule-Thompson expansion \cite{Okcu:2016tgt,Okcu:2017qgo,Mo:2018rgq}.
 Extensions to de Sitter spacetime, in which the pressure becomes a tension \cite{Dolan:2013ft}, have recently been considered \cite{Kubiznak:2015bya,Mbarek:2018bau,Simovic:2018tdy,Simovic:2019zgb}
and even to dynamical cosmological constants  \cite{Gregory:2017sor}.  Accelerating black holes 
have been shown to have a sensible thermodynamic interpretation
\cite{Appels:2016uha,Astorino:2016xiy,Astorino:2016ybm,Appels:2017xoe,Anabalon:2018ydc,Anabalon:2018qfv}, where snap transitions from VdW behaviour to Hawking-Page type behaviour have been seen \cite{Abbasvandi:2018vsh,Abbasvandi:2019vfz,Gregory:2019dtq}. Extensions to spacetimes with NUT charge have also been under
recent consideration  \cite{Johnson:2014xza,Johnson:2017ood} and suggest that perhaps 
that the thermodynamics of Lorentzian black holes with NUT charge can be understood
\cite{Kubiznak:2019yiu,Bordo:2019rhu,Bordo:2019tyh,Ballon:2019uha}. 
 The implications of this are that whatever the microstructure of black holes
 may be, it must be rich enough to account for this broad range of phase behaviour \cite{Wei:2015iwa}.  Indeed,  subtle differences between their microstructures of Van der Waals fluids and charged AdS black holes
have recently been pointed out \cite{Wei:2019uqg}.   

In this paper we explore the thermodynamic behaviour of a charged hairy black hole \cite{Anabalon:2013sra, Anabalon:2015ija} in  the context of black hole chemistry \cite{Kubiznak:2014zwa}.\footnote{The extended thermodynamics was studied for black holes that break the hyperscaling symmetry of AdS in \cite{Pedraza:2018eey,Kastor:2018cqc}.} It was recently shown that the dilaton potential of the theory corresponds to an extended supergravity model with dyonic Fayet-Iliopoulos terms. Therefore, the theory is consistent and has a well defined (stable) ground state. Interestingly, when $\Lambda=0$, these theories  with a non-trivial self-interacting term for the dilaton contain thermodynamically stable asymptotically flat black holes \cite{Astefanesei:2019mds}. Intuitively, the existence of a dilaton potential that behaves as a `box' for the hair living outside the horizon suggests unexpected new features meriting investigation. On the other hand, the AdS solutions can clarify aspects of the AdS/CMT correspondence because they can provide information about distinct holographic phases of matter (see, e.g., \cite{Charmousis:2010zz}). Particularly, some  general results for the speed of sound of the dual field theory were presented in \cite{Anabalon:2016yfg, Anabalon:2017eri}.

The particular case we are interested in is a class of exact hairy AdS black hole solutions of  Einstein-AdS gravity with an electric field non-minimally coupled to a scalar field. Without the dilaton potential, these hairy black hole solutions are singular in the zero temperature limit. However, by using the entropy function formalism \cite{Sen:2005wa, Astefanesei:2006dd}, it was shown \cite{Anabalon:2013sra} that, when the dilaton potential is turned on, the extremal hairy black holes have  a finite horizon area. Therefore, the canonical ensemble is well defined.\footnote{For fixed charge, the thermal AdS is not a solution of the equations of motion, but one can use the extremal black hole as background.} This example could be important when considering quantum phase transitions \cite{Sachdev:2019bjn}. We find that these black holes exhibit  non-trivial critical behaviour in the grand canonical ensemble, quite unlike their hairless RN-AdS counterparts.
 
The remainder of the paper is organized as follows. In Section \ref{sec:charged}, we review the thermodynamics of Reissner-N\"ordstrom-AdS (RN-AdS) black holes in the extended phase space. In Section \ref{sec:Hairy}, we analyze the thermodynamic behaviour in the extended phase space for a four-dimensional exact hairy black hole solution to Einstein-Maxwell-dilaton theory with a non-trivial self-interacting scalar field. Some novel and interesting thermodynamic properties of this solution will be presented, such as double critical point in canonical ensemble, and criticality in grand canonical ensemble.

 
\section{P-V criticality of RN-AdS Black Hole}
\label{sec:charged}
The thermodynamics of electrically charged black holes in AdS is quite rich. For example,  a non-trivial gauge potential allows criticality phenomena when the electric charge $Q$ is kept fixed.  This ensemble is achieved by imposing a particular boundary condition for the gauge field.  The thermodynamics of charged AdS black holes for fixed $\Lambda$  \cite{Chamblin:1999tk,Chamblin:1999hg} 
generalizes and becomes fully analogous to Van der Waals fluids in the   extended phase 
space \cite{Kubiznak:2012wp}.

The first law of a four-dimensional RN-AdS  black hole is
\begin{equation}\label{law1}dE=TdS+\Phi dQ\end{equation}
where $E$ is the total energy of the system, $T$ and $S$ are the Hawking temperature and Bekenstein-Hawking entropy, and $Q$ and $\Phi$ are the electric charge and (conjugate) electrostatic potential. The consideration of a negative cosmological constant, defined as $\Lambda =-3/L^2$ ($L$ is the radius of AdS), as a variable in its own right, playing the role of a pressure $P$ of spacetime as 
\begin{equation}P=-\frac{\Lambda}{\kappa}
\end{equation}
where $\kappa=8\pi G_N$ (with our conventions, we shall set $G_N=c=1$), provides another perspective on the thermodynamic nature of asymptotically AdS black holes.
The proposed first law in the extended phase space, including a dynamical cosmological constant, becomes 
\begin{equation}dE=TdS+\Phi dQ+VdP
\label{first1}
\end{equation}
where
\begin{equation}\label{Vdef}
V\equiv\left(\frac{\pa M}{\pa P}\right)_{S,Q}
\end{equation}
is the `thermodynamic volume' (the thermodynamic quantity conjugate of the pressure). We shall later comment in section \ref{Discuss}   on a different interpretation more suitable in  the context of AdS-CFT duality. This extended form of the first law is compatible with the Smarr relation
\begin{equation}\label{Smarr}
E=2TS+Q\Phi-2PV
\end{equation}
that follows from scaling relations between $E,S,Q$, and $P$ \cite{Kastor:2009wy}.

For completeness and clarity, we review in this section the basic thermodynamic behaviour of RN-AdS black hole, which is a static spherically symmetric solution of the Einstein-Maxwell theory given by the action
\begin{equation}
I_{bulk}[g_{\mu\nu},A_\mu]=\frac{1}{2\kappa}
\int_{\mathcal{M}}{d^4x\sqrt{-g}
	\left(R-F^2-2\Lambda\right)}
\label{action1}
\end{equation}
where $R$ is the Ricci scalar, $F^2\equiv F_{\mu\nu}F^{\mu\nu}$, $F_{\mu\nu}=\pa_{\mu}A_\nu-\pa_{\nu}A_\mu$, and $A_\mu$ is the gauge potential. The corresponding equations of motion,
\begin{align}
R_{\mu\nu}-\frac{1}{2}g_{\mu\nu}R+{\Lambda}g_{\mu\nu}
&=2\left(F_{\mu\alpha}F_\nu{}^{\alpha}-\frac{1}{4}g_{\mu\nu}F^2\right)
\\ \pa_\mu\left(\sqrt{-g}{F}^{\mu\nu}\right)&=0
\end{align}
are solved by
\begin{align}
ds^2&=g_{\mu\nu}dx^\mu dx^\nu
=-f(r){d}t^2+f(r)^{-1}{d}r^2
+r^2d\sigma^2 \label{metric1} \\
A_{\mu}&=\(Q/r+C\)\delta_\mu^t
\label{maxwell1}
\end{align}
where $d\sigma^2\equiv{d}\theta^2+\sin^2\theta\,{d}\varphi^2$ and $C$ is a constant. The metric function is
\begin{equation}
f(r)=
-\frac{\Lambda{r}^2}{3}+1-\frac{2M}{r}+\frac{Q^2}{r^2}
\end{equation}
where $M$ is the mass and $Q$ is the physical electric charge.\footnote{The electric charge is the conserved quantity obtained from the Gauss Law at  spatial infinity $$Q=\frac{1}{4\pi}\oint_{s_\infty^2}{\star F}$$ where $\star F\equiv \(\frac{1}{4}\sqrt{-g}\epsilon_{\alpha\beta\mu\nu}\)F^{\alpha\beta}dx^\mu\wedge dx^\nu$ and $\epsilon_{\alpha\beta\mu\nu}$ is the totally antisymetric Levi-Civita symbol.} Black holes exist provided $f(r_+)=0$, where $r_+$ is the location of the outer event horizon. The additive constant introduced in the gauge potential (\ref{maxwell1}) is chosen to fix $A_t(r_+)=0$. This choice will allow us to properly define the thermodynamic ensembles later on this section. It is more convenient to express the thermodynamic quantities using as parameters the outer horizon radius, $r_+$, and physical charge $Q$, rather than working with the mass $M$:
\begin{equation}
M=\frac{r_+}{2}\(1-\frac{\Lambda{}r_+^2}{3}+\frac{Q^2}{r_+^2}\),\quad
T=\frac{1}{4\pi r_+}
\left(1-\Lambda r_+^2-\frac{Q^2}{r_+^2}\right)
,\quad
S=\pi r_+^2
\label{mass}
\end{equation}
The conjugate potential $\Phi$ is defined as the difference of electrostatic potential between the event horizon and boundary $r\rightarrow\infty$,
\begin{equation}\label{Psi}
\Phi\equiv A_t(r_+)-A_t(\infty)=\frac{Q}{r_+}
\end{equation}
The first law \eqref{first1} and Smarr relation \eqref{Smarr} are satisfied provided $V=\frac{4}{3}\pi r_+^3$, which follows from \eqref{Vdef}.

\subsection{Grand Canonical Ensemble ($\Phi$ fixed)}

\begin{figure}[h]
	\centering
	\subfigure{\includegraphics[width=0.32\textwidth]{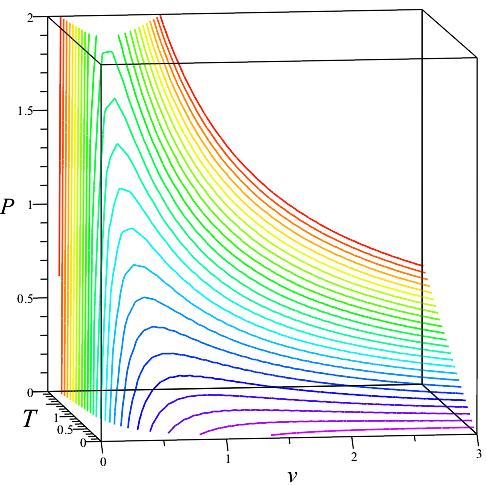}}\quad
	\subfigure{\includegraphics[width=0.32\textwidth]{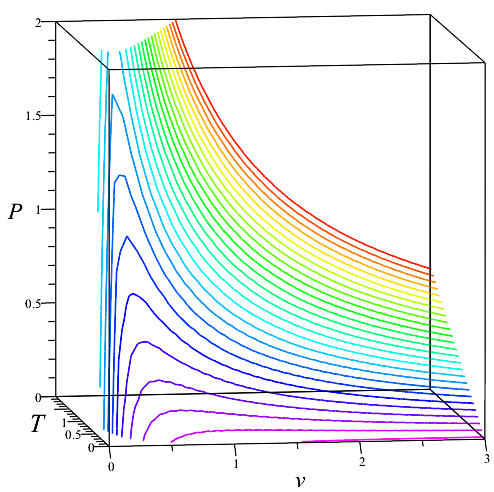}}
	\subfigure{\includegraphics[width=0.32\textwidth]{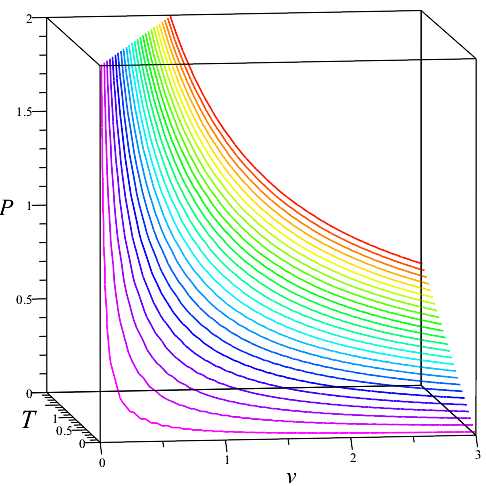}}
	\caption{\small Equation of state for $\Phi=0$ (left), $\Phi=0.8$ (middle) and $\Phi=1$ (right). In the interval $0\leq\Phi<1$, the surface develops a maximum when $T$ is fixed. For $\Phi>1$, the isotherms $P$-$v$ do not develop a local maximum.}
	\label{RN1}
\end{figure}

Let us consider the action (\ref{action1}) supplemented with the Gibbons-Hawking boundary term and  gravitational counterterm \cite{Balasubramanian:1999re,Mann:1999pc} needed to remove the divergences of the action,
\begin{equation}
I=I_{bulk}
+\frac{1}{\kappa}\int_{\pa\mathcal{M}}{d^3x\sqrt{-h}K}
-\frac{1}{\kappa}\int_{\pa\mathcal{M}}{d^3x\sqrt{-h} \left[\frac{2}{L}+\frac{L\mathcal{R}^{(3)}}{2}\right]}
+I_A
\label{counter}
\end{equation}
The trace of extrinsic curvature, $K_{\mu\nu}\equiv\nabla_\mu n_\nu$, is $K= h^{\mu\nu}K_{\mu\nu}$, and $n_\mu$ is the unit normal vector on the $r=$\,const. hypersurface. In (\ref{counter}), $I_A$ is the boundary term associated with the gauge field required for a well posed action principle. For grand canonical ensemble, given by the boundary condition $\left.\delta A_\mu\right|_{\pa\mathcal{M}}=0$, which is the case when $\Phi$ is kept fixed,\footnote{The proof is as follows:
	$\delta A_\mu|_{\pa\mathcal{M}}
	=\left.\delta\(\tfrac{Q}{r}-\tfrac{Q}{r_+}\)
	\right|_{\pa\mathcal{M}}\delta_\mu^t
	=-(\delta\Phi) \delta_\mu^t +\mathcal{O}(r^{-1})=0
	\;\Leftrightarrow\; \delta\Phi=0 \quad$.} it turns out that $I_A=0$. To obtain the canonical ensemble, however, one must add a finite contribution \cite{Hawking:1995ap,Chamblin:1999tk},
\begin{equation}
I_A= \frac{2}{\kappa}\int_{\pa\mathcal{M}}
{d^3x\sqrt{-h}n_\mu F^{\mu\nu} A_\nu}
\label{boundaryA}
\end{equation}
Inserting the solution (\ref{metric1})--(\ref{maxwell1}) into the action (\ref{counter}) on the Euclidean section,\footnote{We use the standard analytic continuation $t\rightarrow -i\tau_E$, where $0<\tau_E<\beta\equiv T^{-1}$.} we obtain the thermodynamic potential $\mathcal{G}$ for the grand canonical ensemble,
\begin{equation}
\mathcal{G}(T,\Phi,P)\equiv I_ E/\beta =\frac{1}{2}(r_{+}-M) =M-TS-Q\Phi 
\end{equation}
Noting that $d\mathcal{G}(T,\Phi,P)=-SdT-Qd\Phi+VdP$, upon using the first law (\ref{first1}) we have
\begin{equation}
S=-\left(\frac{\pa\mathcal{G}}{\pa T}\right)_{\Phi,P}
,\quad
Q=-\left(\frac{\pa\mathcal{G}}{\pa\Phi}\right)_{T,P}
,\quad
V=\left(\frac{\pa\mathcal{G}}{\pa P}\right)_{T,\Phi}
\end{equation}
In order to study the thermodynamic behaviour of this system when $\Phi$ is fixed, let us first obtain the   equation of state. By introducing the `specific volume' $v$, defined by the thermodynamic volume divided by the total number of states (proportional to the entropy)
\cite{Gunasekaran:2012dq}
\begin{equation}
v\equiv \frac{3V}{2S}=2r_+
\label{specificvol}
\end{equation} 
and using it to eliminate $r_+$, we rearrange the equation for temperature, (\ref{mass}), to get
\begin{equation}
T=Pv+\left(\frac{1-\Phi^2}{2\pi}\right)\frac{1}{v}
\label{state1}
\end{equation}
which is a particular case of the Van der Waals  equation of state. 
Figure \ref{RN1} shows the behaviour of the equation of state (\ref{state1}) for different fixed values of $\Phi\in(0,1)$.

\begin{figure}[h]
	\centering
	\subfigure{\includegraphics
		[width=0.3\textwidth]{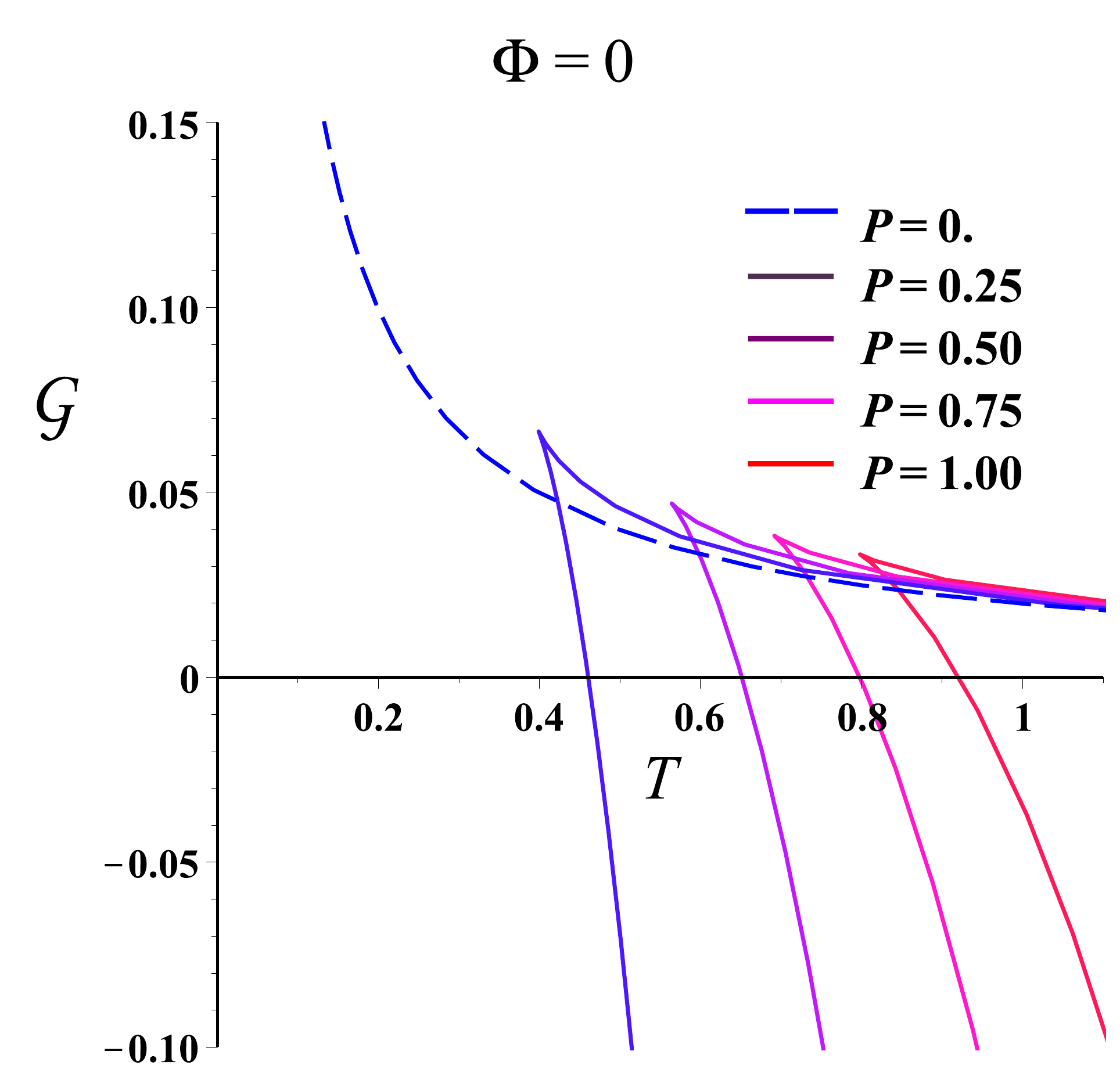}}
	\,\,\,\,\,\,\,\,\,\,\,\,\,\,\,\,\,\,\,\,\,\,
	\,\,\,\,\,\,\,
	\subfigure{\includegraphics
		[width=0.3\textwidth]{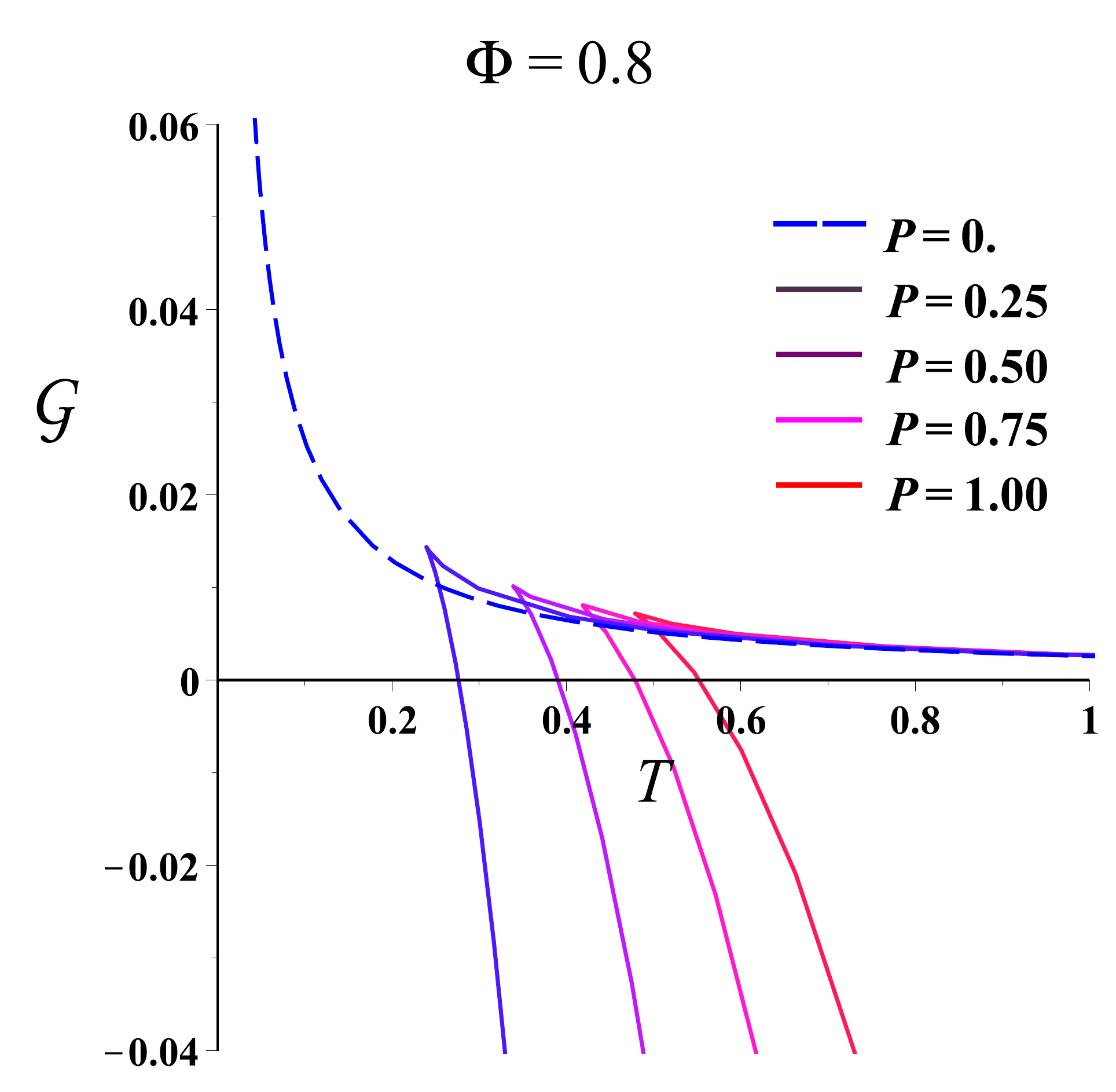}}
	\caption{Thermodynamic potential vs temperature, for $\Phi<1$. In this interval for the conjugate potential, there is a first order (Hawking-Page) phase transition for any $P>0$.}
	\label{fig2}	
\end{figure}

\begin{figure}[h]
\centering
\subfigure{\includegraphics
[width=0.34\textwidth]{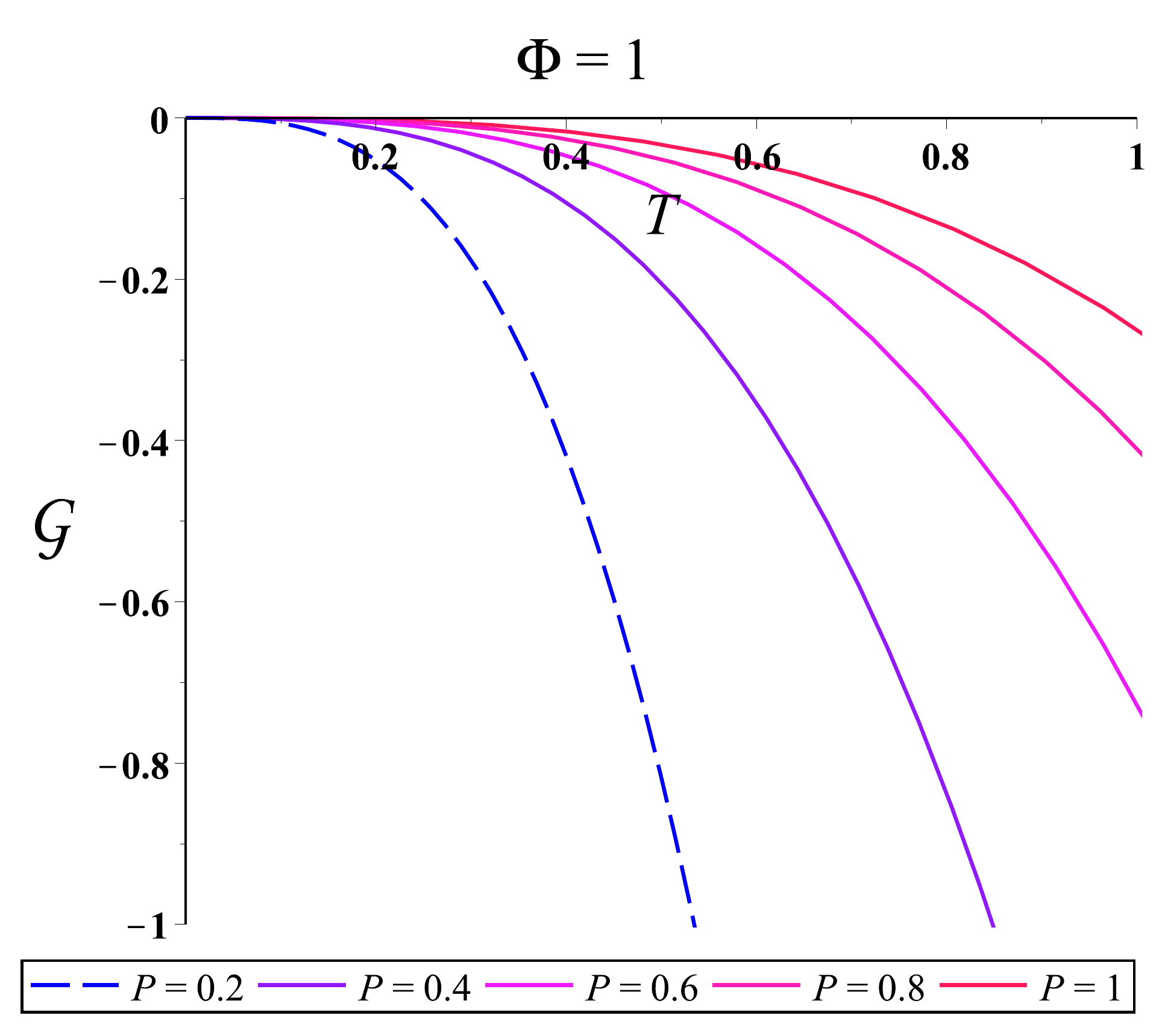}}
\,\,\,\,\,\,\,\,\,\,\,\,\,\,\,\,\,\,\,\,\,\,
\subfigure{\includegraphics
[width=0.34\textwidth]{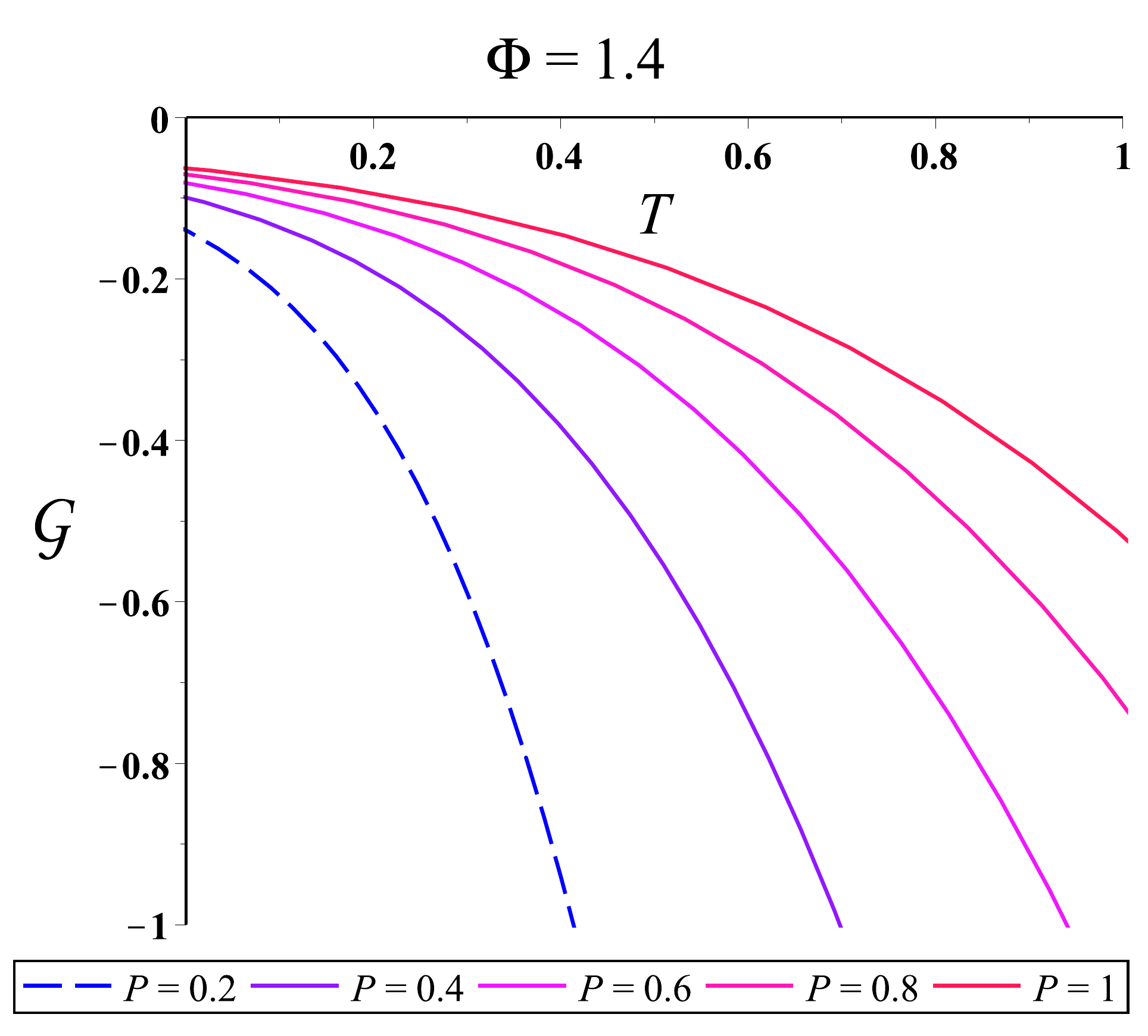}}
\caption{\small Thermodynamic potential vs temperature, for $\Phi\geq 1$. In this interval, there is no Hawking-Page phase transition. The extremal limit, $T=0$, is reached at a finite entropy.}
\label{fig3}	
\end{figure}

In the interval $0\leq \Phi<1$, the surface has a bend and, for a fixed temperature, the pressure develops a maximum $P_{\text{max}}=\frac{\pi T^2}{2\(1-\Phi^2\)}$ at $v=\frac{1-\Phi^2}{\pi T}$, given by the condition $\left({\pa P}/{\pa v}\right)_{T,\Phi}=0$.
However, nowhere in the parameter space does the second derivative $({\pa^2P}/{\pa v^2})_{T,\Phi}=\frac{2P}{v^2}
-\frac{2(1-\Phi^2)}{\pi v^4}$ simultaneously vanish with the first derivative and so there is no critical point in the grand canonical ensemble.
In Figure \ref{fig2} and \ref{fig3},  we depict the thermodynamic potential,
\begin{equation}
\mathcal{G}(r_+,P,\Phi)
=\frac{r_+}{12}\[3\(1-\Phi^2\)-8\pi Pr_+^2\]
\end{equation}
as a function of temperature, for different values of the conjugate potential. It is easy to see that a first-order phase transition 
\cite{Chamblin:1999tk, Chamblin:1999hg,Kubiznak:2014zwa}, similar to the
Hawking-Page transition \cite{Hawking:1982dh} for the neutral black hole, occurs in the interval $0\leq\Phi<1$, where the charged black hole discharges and dissolves into pure radiation with  $\mathcal{G}=0$. The main point is that the thermodynamic ensemble is different --– it is the grand canonical ensemble, defined by coupling the system to energy and charge reservoirs at fixed temperature $T$ and potential $\Phi$. When the black hole undergoes a phase transition it discharges, its charge going into the charge reservoir. This takes place at the temperature
\begin{equation}
T_{HP}=
\[\frac{8\(1-\Phi^2\)P}{3\pi}\]^{1/2}
\label{hp1}
\end{equation}
which is where $\mathcal{G}(T_{HP})=0=\mathcal{G}_{AdS}$. Observe from equation (\ref{hp1}) that, provided $0\leq\Phi<1$, the Hawking-Page phase transition occurs at a finite temperature for any positive pressure, for black holes whose entropy is $S_{HP}=\frac {3(1-\Phi^2)}{8P}$.
Within this range, there exists a `cusp'. At this point, the second derivative of $\mathcal{G}$ is discontinous, while the first derivative of $\mathcal{G}$ remains continuous.

The cusp, that corresponds to the second derivative of the thermodynamic potential at fixed $\Phi$  (or, equivalently, to the discontinuity in the heat capacity), is located at 
\begin{equation}
T_{cusp}
=\sqrt{\frac{2\(1-\Phi^2\)P}{\pi}},
\qquad
\mathcal{G}_{cusp}=\frac{1}{24}
\sqrt{\frac{2\(1-\Phi^2\)^3}{\pi P}}
\end{equation}
when the entropy is $S_{cusp}=\frac{1-\Phi^2}{8P}$.
We can understand the disappearance of the cusp  for $\Phi>1$ as follows: we observe that for $\phi=1$ the Hawking-Page temperature vanishes and so the black hole is extremal. Therefore, there cannot exist a phase transition of first order when $\Phi>1$ and that is confirmed in Figure \ref{fig3}. In those cases, as shown in Figure \ref{fig3}, the extremal limit, $T=0$, is reached for a finite entropy. This can be easily seen by solving $r_+$ from $T=0$ in equation (\ref{mass}) for temperature, and evaluating $S=\pi r_+^2$,
\begin{equation}
T=0\quad
\rightarrow\quad
S_{\text{extremal}}=
\frac {\Phi^2-1}{8P}
\end{equation}
The case $\Phi=1$ is special because the horizon shrinks to a null singularity and so the entropy vanishes, $S_{\text{extremal}}=0$. This corresponds to the supersymetric limit when $M=Q$ and that is why the thermodynamic potential also vanishes, $\mathcal{G}=0$ in this particular case.

\begin{figure}[h]\centering	\subfigure{\includegraphics		[width=0.42\textwidth]{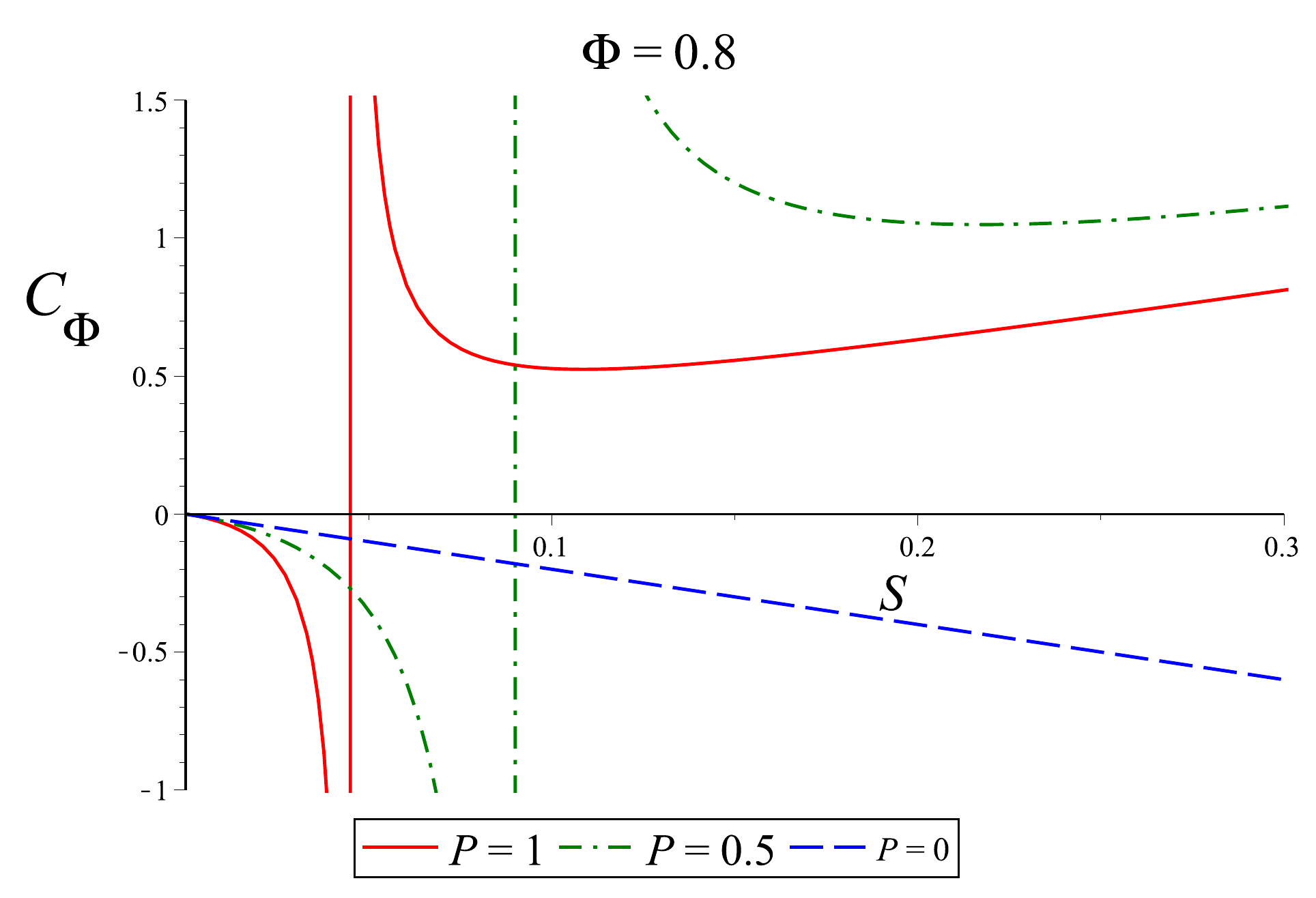}}	\,\,\,\,\,	\subfigure{\includegraphics		[width=0.42\textwidth]{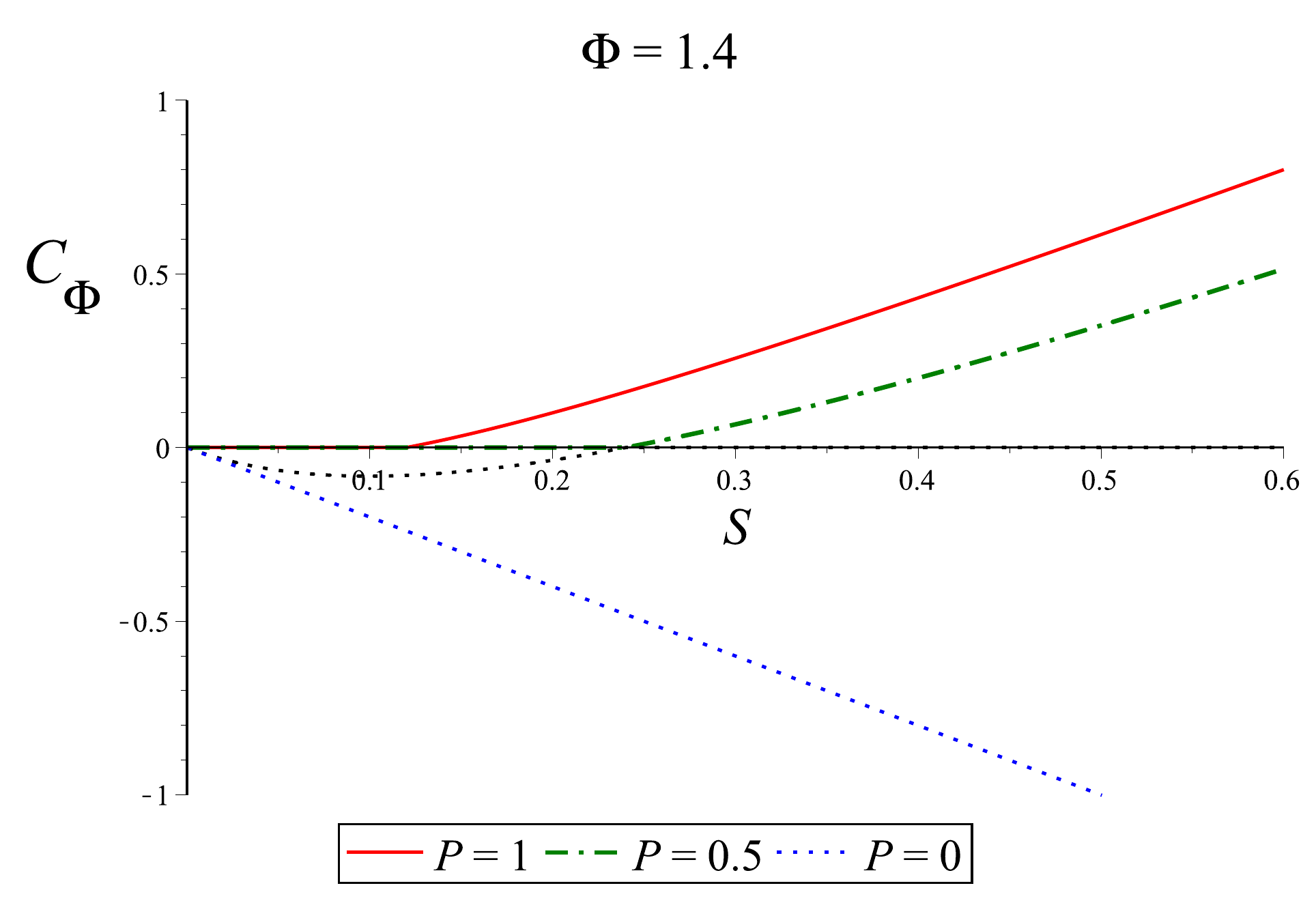}}	\caption{\small Heat capacity as a function of the entropy, for $\Phi=0.8$ and $\Phi=1.4$. In the first case, representing $0<\Phi<1$, we observe small thermally unstable and large thermally stable black holes. For the second case, representing $\Phi>1$, the black holes become thermally stable, $C_\Phi>0$.  In the second plot, the continuation in dotted lines indicates unphysical regions for which $T<0$. }
\label{heat1}
\end{figure}

The response functions are important in analyzing   thermodynamic stability. The heat capacity $C_\Phi\equiv T\(\pa S/\pa T\)_\Phi$ and the electric permittivity $\epsilon_S\equiv\(\pa Q/\pa\Phi\)_S$ are relevant for stability in the grand canonical ensemble, provided $P$ is fixed. Note that, for $\Phi<1$, there are two black hole configurations at a given temperature, corresponding to a small and a large black hole. On the other hand, for $\Phi\geq 1$, there is only one configuration.
It is easy to compute the permittivity as $\epsilon_S=r_+>0$, which is the same for any configuration. On the other hand, since $C_\Phi=-T\(\pa\mathcal{G}^2/\pa T^2\)_{\Phi,P}$ and $S=-\({\pa\mathcal{G}/\pa T}\)_{\Phi,P}$, only by observing the slope and concavity of the thermodynamic potential for $0\leq\Phi<1$, we can conclude that the black holes with $S<S_{cusp}$ are thermally unstable and with $S>S_{cusp}$, thermally stable. For $\Phi\geq 1$, the cusp disappears and the black holes become thermally stable. This can be explicitly seen in Figure \ref{heat1}.

In Section \ref{sec:Hairy}, we shall compare these results with the ones obtained for the hairy charged black holes in the grand canonical ensemble.

\subsection{Canonical ensemble ($Q$ fixed)}

\begin{figure}[ht]
\centering
\subfigure{\includegraphics
[width=0.4\textwidth]{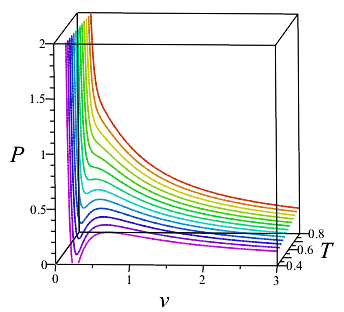}}
\,\,\,\,\,\,\,\,\,\,\,\,\,\,
\subfigure{\includegraphics
[width=0.4\textwidth]{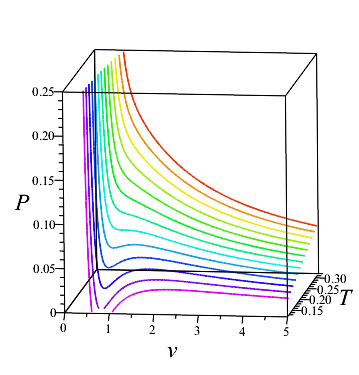}}
\caption{\small Equation of state depicted for $Q=0.1$ (at the left hand side) and $Q=0.3$ (right hand side).}
\label{eqstateRN2}
\end{figure}

By adding the boundary term (\ref{boundaryA}) to the action, we get the thermodynamic potential for canonical ensemble --- this corresponds to a Legendre transform of the thermodynamic potential. On the Euclidean section, this boundary term provides a new contribution,
\begin{equation}
I_A^E=\beta Q\Phi
\end{equation}
The on-shell action for the canonical ensemble becomes then 
\begin{equation}
I^{on-shell}_E=\frac{\beta}{2}\(r_{+}-M+\frac{2Q^2}{r_+}\)=\beta(M-TS)
\end{equation}
and so the new thermodynamic potential is
\begin{equation}
\mathcal{F}(T,Q,P)\equiv I_ E/\beta=M-TS
\end{equation}
or, equivalently, in the differential form: $d\mathcal{F}(T,Q,P)=-SdT+\Phi{d}Q+VdP$.  The following relations are then automatically satisfied
\begin{equation}
S=-\left(\frac{\pa{\mathcal{F}}}{\pa T}\right)_{Q,P}
{~},\quad
\Phi=\left(\frac{\pa{\mathcal{F}}}{\pa Q}\right)_{T,P}
{~},\quad
V=\left(\frac{\pa{\mathcal{F}}}{\pa P}\right)_{T,Q}
\end{equation}
As in the previous subsection, we can get the  equation of state \cite{Kubiznak:2012wp},
\begin{equation}
T=Pv
+\frac{1}{2\pi v}-\frac{2 Q^2}{\pi v^3}
\end{equation}
depicted in Figure \ref{eqstateRN2}.
In contrast with what was obtained in the grand canonical ensemble (\ref{state1}), there is now a term proportional to $v^{-3}$ (purely due to the charge), which allows for a critical point satisfying
\begin{equation}
\(\frac{\pa P}{\pa v}\)_{T,Q}=0, \qquad
\(\frac{\pa^2P}{\pa v^2}\)_{T,Q}=0
\label{criticality1}
\end{equation}
Equations (\ref{criticality1}) can be analytically solved, obtaining the critical point characterized by
\begin{equation}
P_c=\frac{1}{96\pi Q^2}, \qquad 
v_c=2\sqrt{6} |Q|, \qquad 
T_c=\frac{\sqrt{6}}{18\pi |Q|}
\label{crRN}
\end{equation}
These critical values satisfy 
$P_cv_c/T_c=3/8$ \cite{Kubiznak:2012wp}, which is exactly the same value of the Van der Waals fluid. It is interesting that the value $3/8$ does not depend on the charge of the black hole. 
The critical compresibility factor $Z_c\equiv P_cv_c/T_c$ is a basic parameter used for the definition of thermodynamic similarity class of different substances. We shall compare this result with $Z_c$ for hairy black holes, in Section \ref{sec:Hairy}.

\begin{figure}[h]
\centering
\subfigure{\includegraphics
[width=0.6\textwidth]{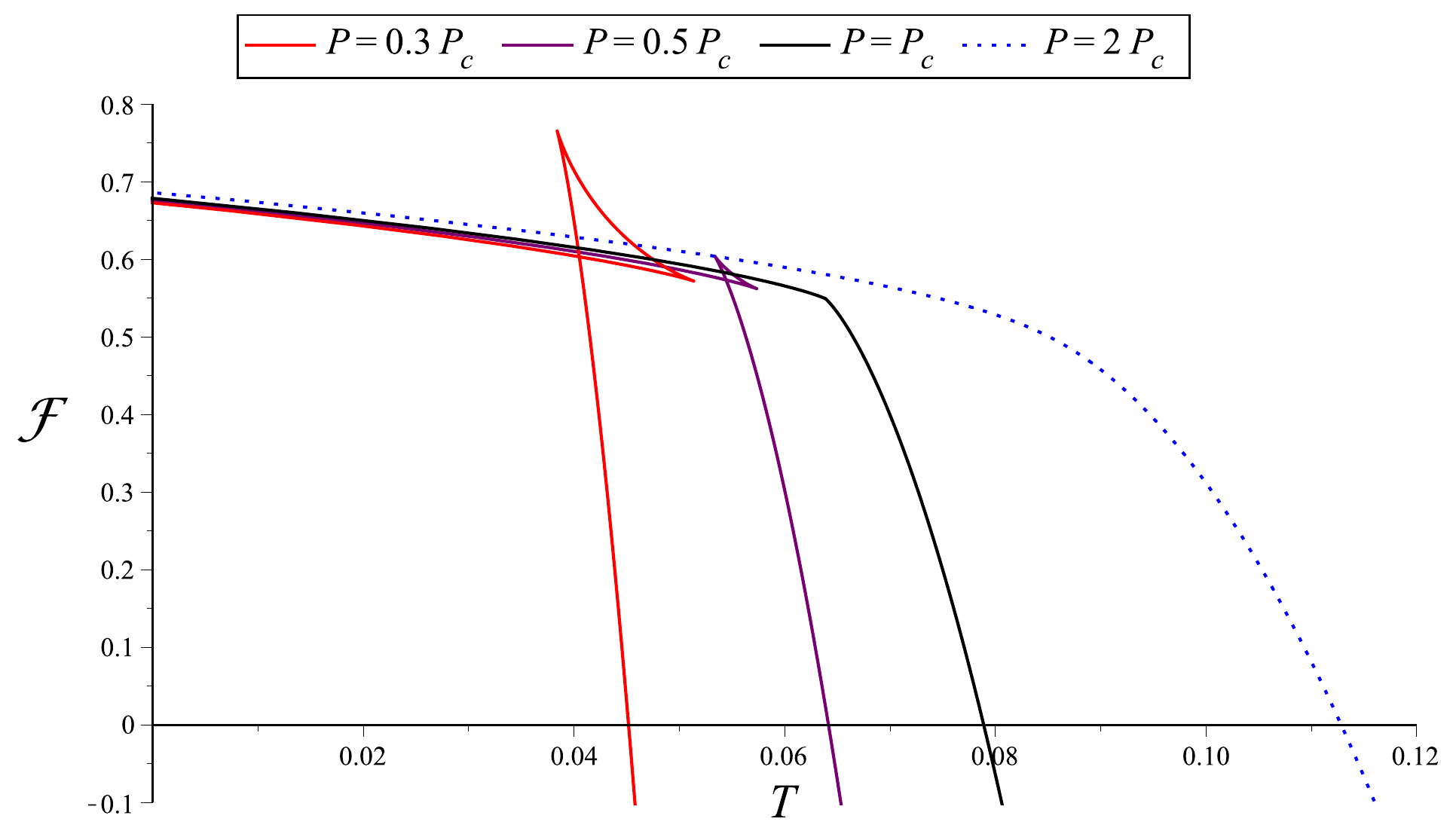}}
\caption{\small Thermodynamic potential in canonical ensemble, for $Q=0.67$. On the critical isobar $P_c\approx 0.0074$, the critical temperature is the temperature where the swallowtail ends, $T_c\approx 0.064$, at $\mathcal{F}_c\approx 0.547$. There is a small-large black hole first-order phase transition for $T<T_c$.}
\label{FQRN}
\end{figure}

The thermodynamic potential that can be parametrically put in the form
\begin{equation}
\mathcal{F}(r_+,Q,P)=
\frac{r_+}{4}\(1+\frac{3Q^2}{r_+^2}-\frac{8\pi Pr_+^2}{3}\)
\end{equation}
is depicted as a function of the temperature in Figure \ref{FQRN}.
The critical point occurs at $\mathcal{F}_c=\frac{\sqrt6}{3} |Q|>0$.

Note that there is no Hawking-Page-like phase transition for fixed charge.  This is because the charge is conserved and so, when $\mathcal{F}=0$, the system cannot dissolve into  pure radiation as this would not conserve charge.  Instead there is a (small black hole)-(large black hole) first-order phase transition at the swallowtail intersection.


\section{Hairy electrically charged AdS black hole}
\label{sec:Hairy}
\begin{figure}[h]
	\centering
	\subfigure{\includegraphics
		[width=0.75\textwidth]{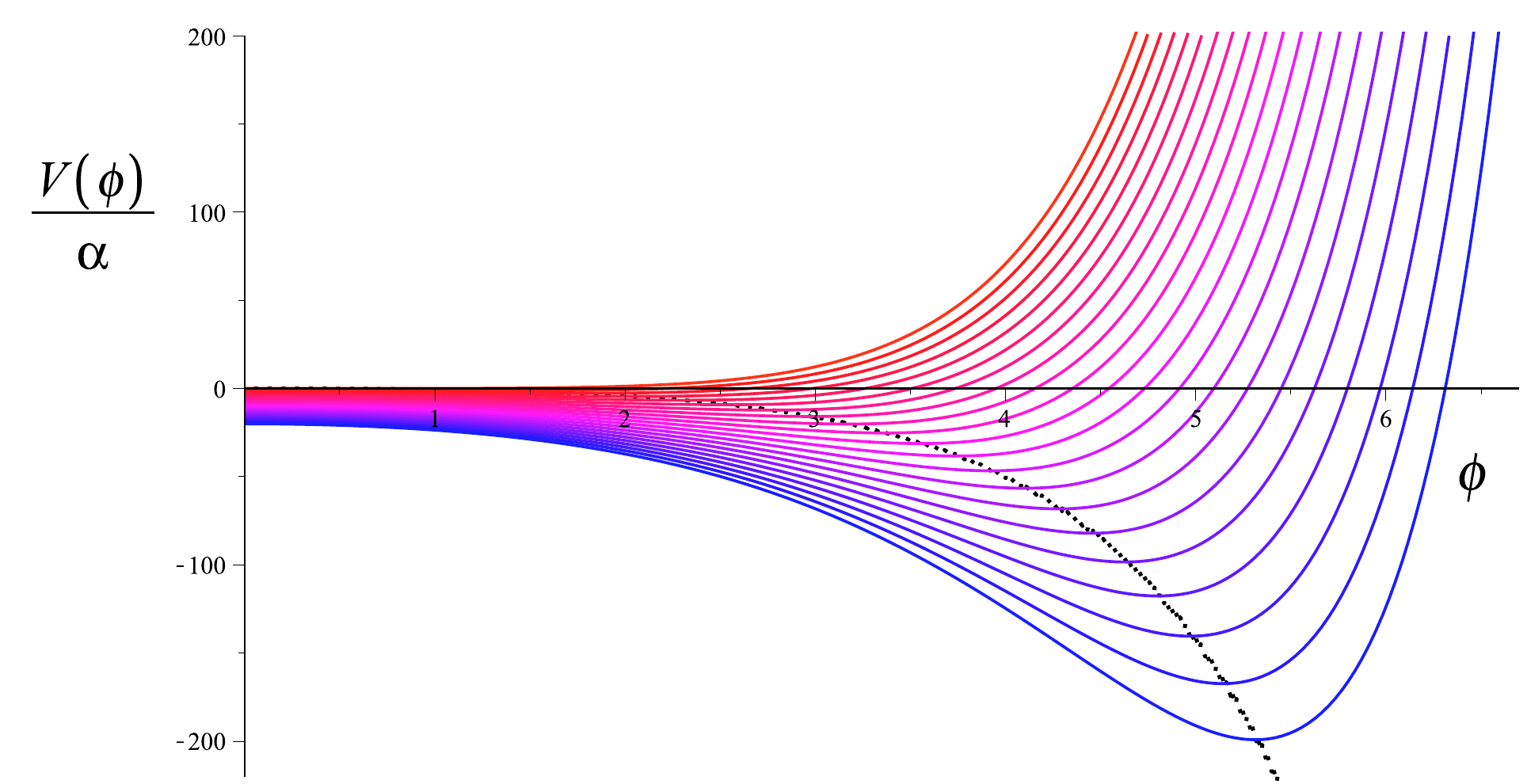}}
	\caption{\small The dilaton potential (over $\alpha$) vs $\phi$. Red curve at the top of the figure is the case $\Lambda=0$, while the curve in blue, at the bottom, corresponds to $\alpha^{-1}\Lambda=-10$. Since $V$ has the same dimension as $\alpha$ and $\Lambda$, then $\alpha^{-1}V$ and $\alpha^{-1}\Lambda$ are both dimensionless.}
	\label{dilatonpot}
\end{figure}

In this section, we analyze in detail the thermodynamics and phase transitions for exact regular hairy black hole solutions in a theory with the action
\begin{equation}
I\[g_{\mu\nu},A_\mu,\phi\]
=\frac{1}{2\kappa}\int_{\mathcal{M}}{d^4x\sqrt{-g}
\[R-e^{\phi}F^2-\frac{1}{2}\(\pa\phi\)^2-V(\phi)\]}
\label{action2}
\end{equation}
$V(\phi)$ is the self-interaction potential and the scalar field is non minimally coupled to the gauge field. The equations of motion are
\begin{align}
R_{\mu\nu}-\frac{1}{2}g_{\mu\nu}R
&=T_{\mu\nu}^{EM}+T_{\mu\nu}^\phi\\
\pa_{\mu}
\left(\sqrt{-g}e^{\phi}F^{\mu\nu}\right)&=0\\
\frac{1}{\sqrt{-g}}\pa_{\mu}
\left(\sqrt{-g}g^{\mu\nu}\pa_{\nu}\phi\right)
&=\frac{dV(\phi)}{d\phi}
+e^{\phi}F^2
\end{align}
where the energy-momentum tensors for the gauge potential and dilaton are
\begin{equation}
T_{\mu\nu}^{A}=2e^{\phi}
\(F_{\mu\alpha}F_{\nu}{}^{\alpha}-\frac{1}{4}\,g_{\mu\nu}F^2\),
\quad
T_{\mu\nu}^{\phi}
=\frac{1}{2}\,\pa_{\mu}\phi\,\pa_{\nu}\phi
-\frac{1}{2}g_{\mu\nu}\[\frac{1}{2}\(\pa\phi\)^2+V(\phi)\]
\end{equation}
The solution we are going to analyze was presented in \cite{Anabalon:2013sra} for asymptotically AdS spacetime\footnote{The asymptotically flat solutions can be obtained in the limit for which $\Lambda$ vanishes, though the potential is still non-trivial in this case \cite{Anabalon:2013qua,Astefanesei:2019mds}. Different asymptotically flat solutions that can support scalar hair were recently reported in \cite{Hong:2019mcj}.} as a particular case of a more general class of exact solutions \cite{Acena:2013jya,Anabalon:2013qua,Anabalon:2015vda,Acena:2012mr,Anabalon:2013eaa,Anabalon:2016izw}. The equations of motion can be analytically solved with the following dilaton potential
\begin{equation}
\label{dilaton}
V(\phi)=\(\frac{\Lambda}{3}+\alpha\phi\)(4+2\cosh\phi)-6\alpha\sinh\phi
\end{equation}
where $\alpha$ is a parameter which is related to the dyonic Fayet-Iliopoulos SUGRA sector \cite{Anabalon:2017yhv} and $\Lambda$ is the cosmological constant, assumed to be negative. Note that $V(0)=2\Lambda$, consistent with the $AdS$ asymptotics:
\begin{equation}
\left.\frac{dV}{d\phi}\right|_{\phi=0}=0 \,, \qquad \left.\frac{d^2V}{d\phi^2}\right|_{\phi=0}=\frac{2\Lambda}{3}
\end{equation}
The potential's behaviour is shown in Figure \ref{dilatonpot}, where we observe the existence of a global minimum.

\begin{figure}[h]
\centering
\subfigure{\includegraphics
[width=0.60\textwidth]{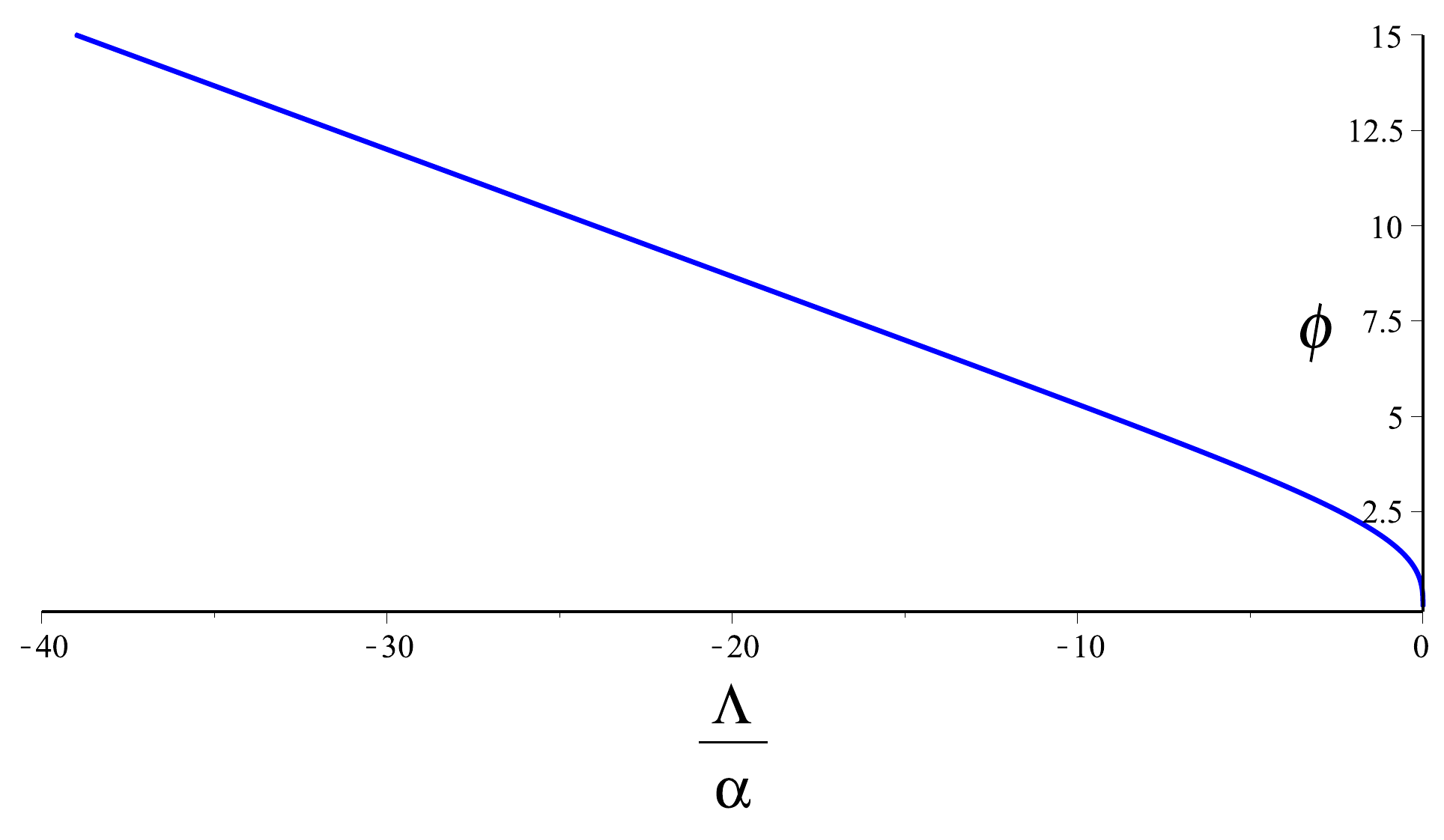}}
\caption{\small The curve shows the points where the first derivative $dV/d\phi$ vanishes. For any $\Lambda$ the potential develops a global minimum for some $\phi>0$.}
\label{dilatonpot2}
\end{figure}

In Figure \ref{dilatonpot2}, we  depict the curve along which the first derivative of the dilaton potential vanishes, which is given by the implicit equation
\begin{equation}
2\(\frac{\Lambda}{3}+\alpha\phi\)\sinh\phi
+4\alpha(1-\cosh\phi)
=0
\end{equation}
where we can observe that, for a fixed given $\alpha$, there is a value of the scalar field at which the potential has a global minimum. This is true for any $\Lambda<0$.


\newpage
The metric, gauge potential, and dilaton are given by
\begin{align}
\label{metric}
ds^2&=\Omega(x)
\[-f(x)dt^2+\frac{\eta^2dx^2}{x^2f(x)}+
d\theta^2+\sin^2\theta d\phi^2\] \\
A_\mu&=\left(\frac{q}{x}-C\right)\delta^t_\mu\,, \qquad \phi=\ln(x) \label{gauge}
\end{align}
where $C$ is an additive constant and the explicit expressions for the conformal factor $\Omega(x)$ and metric function $f(x)$ are 
\begin{align}
\Omega(x)&=\frac{x}{\eta^2(x-1)^2},\;\;\;
f(x)=\alpha\(\frac{x^2-1}{2x}-\ln{x}\)
+\frac{\eta^2(x-1)^2}{x}\(1-\frac{2q^2}{x}\)
-\frac{\Lambda}{3}
\label{conformal}
\end{align}
Black holes exist provided the horizon equation $f(x_+)=0$ is satisfied. The constant $C$ in the gauge potential is again chosen to fix $A_t(x_+)=0$.
Without loss of generality, because it appears as $\eta^2$ in the solution, the integration constant $\eta$ is chosen to be a positive definite quantity. In this coordinate system, the solution is known for describing two disconnected spacetimes, since one can approach   the boundary (located at $x\to1$) either from the `left' ($x\rightarrow 1^{-}$) or from the `right' ($x\rightarrow 1^{+}$). The negative branch is defined by the domain $0<x<1$; for this spacetime, $\phi< 0$ (hence the name), whereas the positive branch corresponds to the range $1< x <\infty$ when the scalar field is positive definite (for more details on the general solution and its properties, see \cite{Anabalon:2013sra}). We emphasize that the scalar field and Ricci scalar diverge at $x = \infty$ and so this point corresponds to the singularity of the hairy black hole.

The thermodynamic properties of these two branches must be independently studied.  We will shall focus only on the positive branch that, interestingly,  in the limit $\Lambda=0$  corresponds to thermodynamically stable black holes in flat space \cite{Astefanesei:2019mds}.


\subsection{Counterterm method and the usual thermodynamics}
The counterterm method in AdS was developed in \cite{Balasubramanian:1999re,Skenderis:2000in,Henningson:1998gx, Mann:1999pc, Emparan:1999pm} and, in the presence of scalar fields with mixed boundary condition, in \cite{Marolf:2006nd, Papadimitriou:2007sj, Anabalon:2015xvl}. This method was successfully applied not only to black hole solutions, but also to other solutions that are locally asymptotically AdS \cite{Balasubramanian:2002am, Astefanesei:2005yj,Astefanesei:2005eq,Astefanesei:2004kn,Astefanesei:2004ji, Balasubramanian:2005bg}.

In this subsection, we are going to compute the on-shell Euclidean action and show that  the quantum statistical relation is satisfied. In this regard, we should consider the full regularized action that consists of the bulk part, Gibbons-Hawking boundary term $I_{GH}$, gravitational counterterm for asymptotically AdS spacetime $I_{g}$ \cite{Balasubramanian:1999re}, and a boundary term for the scalar field $I_\phi$ \cite{Astefanesei:2008wz,Mann:2009id,Anabalon:2015xvl}, 
\begin{equation}
I=\frac{1}{2\kappa}\int_{\mathcal{M}}{d^4x\sqrt{-g}
	\[R-e^{\phi}F^2-\frac{1}{2}\(\pa\phi\)^2-V(\phi)\]}
+\frac{1}{\kappa}
\int_{\pa\mathcal{M}}{d^3x\sqrt{-h}K}
+I_g
+I_\phi
\end{equation}
where
\begin{align}
I_g&=-\frac{1}{\kappa}
\int_{\pa\mathcal{M}}{d^3x\sqrt{-h} 
	\[\frac{2}{L}+\frac{L\mathcal{R}^{(3)}}{2}\]}
\\
I_\phi&=-\frac{1}{2\kappa}\int_{\pa\mathcal{M}}
{d^3x\sqrt{-h}\[\,\frac{\phi^2}{2L}
+\frac{W(a)}{La^3}\phi^3\]}
\end{align}
Here, $a$ and $b\equiv dW(a)/da$ are the leading and sub-leading components of the asymptotic expansion of the scalar field in canonical coordinates, $\phi=a/r+b/r^2+\mathcal{O}(1/r^{3})$. With the coordinate change $r=\sqrt{\Omega(x)}$ in the asymptotic limit, the scalar field (\ref{gauge}) has the fall-off $\phi=-\frac{1}{\eta r}+\mathcal{O}(r^{-3})$, that is, $a=-\frac{1}{\eta}$ and $b=0$ (corresponding to $W=0$).


Let us now compute the on-shell action for the grand canonical ensemble, that is for the following boundary condition of the gauge field: $\left.\delta A_\mu\right|_{\pa\mathcal{M}}=0$. 
In order to obtain the on-shell action in canonical ensemble, the boundary term for the gauge field that must be added to the action is, in this case,
\begin{equation}
\label{boundary2}
I_A=\frac{2}{\kappa}\int_{\pa\mathcal{M}}{d^3x\sqrt{-h}e^{\phi}n_\mu F^{\mu\nu}A_{\nu}}
\end{equation}
The total (regularized) action is
\begin{align}
I^E&=I_{bulk}^{E}+I_{GH}^{E}+I_\phi^E+I_g^E \\
&=\frac{\beta}{2\eta(x_{+}-1)}
-\frac{\beta}{2\eta(x-1)}
+\frac{\beta}{4\eta}\left.
\(xf'\Omega+2xf\Omega'\)\right|_{x_b}\\
&\qquad\qquad\qquad\qquad\qquad
+\frac{\beta}{L}\left.\[f^{1/2}\Omega^{3/2}
\(1+\frac{\phi^2}{8}\)\]\right|_{x_b}
+\frac{\beta L}{2}\left.\(
f^{1/2}\Omega^{1/2}\)\right|_{x_b}
\notag
\end{align}
and, by evaluating on the solution, we get
\begin{equation}
I^E=\beta\[\frac{q^2}{2\eta}
-\frac{x_{+}+1}{4\eta(x_{+}-1)}
-\frac {\alpha}{24\eta^3}\]+\mathcal{O}\(x_{b}-1\)
\label{onshell1}
\end{equation}

Now, to get the energy of hairy black hole, let us use the quasilocal formalism of Brown and York \cite{Brown:1992br} to obtain the boundary stress tensor, $\tau_{ab}$, defined as 
\begin{equation}
\tau_{ab}\equiv \frac{2}{\sqrt{-g}}\frac{\delta I}{\delta h^{ab}} = -\frac{1}{\kappa}
\(K_{ab}-h_{ab}K+\frac{2}{L}h_{ab}-LG_{ab}\)
-\frac{h_{ab}}{2\kappa L}
\(\frac{\phi^2}{2}+\frac{W(a)}{a^3}\phi^3 \)
\end{equation}
The components of the boundary stress tensor are
\begin{align}
\tau_{tt}&=\frac{1}{\kappa}\[
-\frac{xf^{3/2}\Omega'}{\eta\Omega^{1/2}}
+Lf
+\frac{2\Omega f}{L}\(1+\frac{\phi^2}{8}\)
\]
=\frac{12\eta^2q^2-\alpha}{48\pi\eta^2 L}\,(x_b-1)+\mathcal{O}\[\(x_b-1\)^2\]
\\
\tau_{\theta\theta}&=
\frac{1}{\kappa}\[
-\frac{x(\Omega f'+2f\Omega')}
{2\eta\Omega^{1/2}f^{1/2}}
-\frac{2\Omega}{L}\(1+\frac{\phi^2}{8}\)
\]
=\frac{\(12\eta^2q^2-\alpha\)L}{96\pi\eta^2}\,(x_b-1)+\mathcal{O}\[\(x_b-1\)^2\]
\end{align}
and $\tau_{\varphi\varphi}=\tau_{\theta\theta}\sin^2\theta$.
To obtain the stress tensor of the dual field theory, one has to rescale the boundary stress tensor with the conformal factor $1/(x_b-1)$ \cite{Myers:1999psa}
to get a finite 
expression. One can check, as expected, that the stress tensor is covariantly conserved and its trace vanishes.

The energy of the spacetime is then
\begin{equation}
E
=\oint_{s_\infty^2}{d^2\sigma\sqrt{\sigma}
n^t\tau_{tt}\xi^t}=\frac{12\eta^2q^2-\alpha}{12\eta^3}
\end{equation}
where $\xi=\pa/\pa t$ is the Killing vector and the normal unit to the hypersurface $t=const$ is $n_a=\delta_a^{t}/\sqrt{-g^{tt}}$. We can also obtain the electric charge from the Gauss law at infinity,
\begin{equation}
Q=\frac{1}{4\pi}\oint_{s^2_\infty}{e^\phi\star F}
=\frac{1}{4\pi}
\oint_{s^2_\infty}{\sqrt{-g}e^\phi F^{tx}d\theta\wedge d\phi}
=-\frac{q}{\eta}
\label{charg}
\end{equation}
and its conjugate potential is 
\begin{equation}
\Phi\equiv A_t(x_+)-A_t(x=1)=\frac{q(1-x_+)}{x_+}
\label{potent1}
\end{equation}
By eliminating the conical singularity on the Euclidean section, the black hole's temperature is related to the periodicity of the Euclidean time:
\begin{equation}
T=\frac{1}{\beta}=-\frac{x_+}{4\pi\eta}
\left.\frac{df(x)}{dx}\right|_{x=x_+}
=\frac{(x_{+}-1)^2}{8\pi \eta x_+}
\[-\alpha-2\eta^2\(\frac{x_{+}+1}{x_{+}-1}\)
+4\eta^2q^2\(\frac{x_{+}+2}{x_+}\)\] 
\label{temperh} 
\end{equation}
Finally, with the black hole entropy given by the event horizon area
\begin{equation}
S=\frac{A}{4}=\frac{\pi x_+}{\eta^2(x_{+}-1)^2}
\label{entro}
\end{equation}
one can check that the quantum statistical relation, $\mathcal{G}=I^E/\beta\equiv E-TS-Q\Phi$, and the first law of thermodynamics, $dE=TdS+\Phi dQ$, are satisfied.
\subsection{Extended thermodynamics}

In the previous subsection, we have obtained the thermodynamic quantities for the hairy black hole (\ref{metric}). Since the conformal symmetry is preserved in the boundary, our result for the energy should match the Ashtekar-Magnon-Das (AMD) mass \cite{Ashtekar:1984zz,Ashtekar:1999jx}. Since from a technical point of view it is direct and simple, we take advantage of the method proposed in \cite{Anabalon:2014fla} for general mixed boundary conditions of the scalar field to obtain the AMD mass. Expanding the metric and gauge potential near the boundary and performing the change of coordinates 
$
\Omega(x)=r^2+\mathcal{O}\left(r^{-2}\right)$, we find
\begin{equation}
x=1+\frac{1}{\eta r}+\frac{1}{2\eta^2 r^2}
+\frac{1}{8\eta^3 r^3}+\mathcal{O}\(r^{-4}\)
\label{change1}
\end{equation}
Under this change of coordinates, note that the time component of the gauge potential (\ref{gauge}) becomes $A_t=-\frac{q}{\eta r}+{c_0}+\mathcal{O}\(\frac{1}{r^2}\)$ and so the physical electric charge is $Q=-\tfrac{q}{\eta}$ that matches the result in the previous section.  The mass can be read off from 
\begin{equation}
g_{tt}=1+\frac{\alpha-12\eta^2q^2}{6\eta^3r}
-\frac{1}{3}\Lambda r^2+\frac{q^2}{\eta^2r^2}
+\mathcal{O}\(\frac{1}{r^3}\) \label{ADM1}
\end{equation}
and it consistently follows that the AMD mas matches the energy of the system computed previously by using the quasilocal formalism:
\begin{equation}
M=\frac{12q^2\eta^2-\alpha}{12\eta^3}
\end{equation}
Note also that, from the horizon equation, the integration constant, $\eta$, can be rewritten as a function of the horizon location in the following compact form:
\begin{equation}
\label{horizon}
\eta^2
=
\frac
{3\alpha x_+\(x_+^2-1-2x_+\ln{x_+}\)-2\Lambda x_+^3}
{\[2q^2\(x_{+}-1\)-x_+\](x_+-1)^2}
\end{equation}

Using the dimensional analysis, and supplementing with equation (\ref{horizon}), one can obtain a generalized Smarr formula as
\begin{equation}
M=2TS+\Phi Q-2\(\frac{\pa M}{\pa \Lambda}\)\Lambda
-2\(\frac{\pa M}{\pa\alpha}\)\alpha
\equiv
2TS+\Phi Q-2\mathcal{V}\Lambda-2\mathcal{\omega}\alpha
\label{smarr}
\end{equation}
where $\mathcal{V}\equiv \(\frac{\pa M}{\pa \Lambda}\)_{S,Q,\alpha}$ and $\omega\equiv \(\frac{\pa M}{\pa \alpha}\)_{S,Q,\Lambda}$ have the following expressions:
\begin{align}
\mathcal{V}=-\frac{x_+}{12\eta^3}
\frac{x_{+}+1}{(x_{+}-1)^3}, \quad
{\omega}
=\frac{1}{4\eta^3}
\[\frac{x_+^2+10{x_+}+1}{6\({x_+}-1\)^2}
-\frac{x_+\ln{x_+}\(x_{+}+1\)}{\(x_{+}-1\)^3}\]
\end{align}
One can additionally show that, if $\alpha$ and $\Lambda$ were allowed to represent thermodynamic variables, the first law should be extended by including their variations
\begin{equation}
dM=TdS+\Phi dQ+\mathcal{V}\,d\Lambda+\omega\,d\alpha
\end{equation}
In this work, however, we shall be interested in theories with $\alpha$ fixed and the cosmological constant is treated as the pressure of spacetime, $P=-\frac{\Lambda}{8\pi}$. The thermodynamic volume for this theory, recognized as $V=-8\pi\mathcal{V}$, is
\begin{equation}
{V}\equiv \(\frac{\pa M}{\pa P}\)_{S,Q}
=\frac{2\pi x_+}{3\eta^3}\frac{x_{+}+1}{(x_{+}-1)^3}
\end{equation}
Note that by using the change of coordinate (\ref{change1}), evaluated on the horizon, the thermodynamic volume recieves the first leading contribution from the Euclidean horizon volume plus the hairy corrections
\begin{equation}
{V}=\frac{4\pi r_+^3}{3}+\frac{\pi r_+}{6\eta^2}+\mathcal{O}\(\frac{1}{r_+}\)
\end{equation}

\subsubsection*{Reverse Isoperimetric Inequality}

It has been conjectured that the thermodynamic volume of black holes satisfies the so-called reverse isoperimetric inequality \cite{Cvetic:2010jb}, which states that 
\begin{equation}
\label{ratio}
\mathcal{R}\equiv \[\frac{(d-1)V}{\Omega_{d-2}}\]^{\frac{1}{d-1}}
\(\frac{\Omega_{d-2}}{A}\)^{\frac{1}{d-2}}\geq 1
\end{equation}
where $d$ is the dimensionality of spacetime and $\Omega_{d-2}$ the area of the unit-($d-2$) cross section. 
For RN-AdS, it is easy to check that, with $d=4$ and $\Omega_2=4\pi$ the area of the unit 2-sphere,
\begin{equation}
\mathcal{R}_{RN-AdS}
=\(\frac{3V}{4\pi}\)^{\frac{1}{3}}
\(\frac{4\pi}{A}\)
=1
\end{equation}
since $V=4\pi r_+^3/3$ and $A=4\pi r_+^2$. This can can be interpreted as if, for a given thermodynamic volume, RN-AdS maximized the required entropy.

It can be explicitly proven that, for the hairy black hole under consideration,
the isoperimetric ratio (\ref{ratio}) depends on the horizon location,
\begin{equation}
\mathcal{R}_{hairy}
=\frac{x_{+}+1}{2\sqrt{x_+}} \geq 1
\end{equation}
and satisfies the inequality. Therefore, the reverse isoperimetric inequality remains valid for hairy black holes. In this case, it can be said that, for a fixed thermodynamic volume, the hairy black hole has a smaller entropy than that of the RN-AdS black hole.

To begin the thermodynamic analysis, it will be highly convenient to replace thermodynamic quantities by equivalent dimensionless ones. The rescaling is made by using powers of $\alpha$, which is going to be assumed fixed and positive for this analysis. Then, we will use $\bar\eta=\frac{\eta}{\sqrt{\alpha}}$ and $\bar\Lambda=\frac{\Lambda}{\alpha}$, where the bar symbol ``$\bar{\;\;\;}$'' denotes dimensionless quantities. The dimensionless version of the thermodynamic quantities are
\begin{equation}
\bar M=\sqrt\alpha{M}, \quad
\bar  T=\frac{T}{\sqrt\alpha}, \quad
\bar {S}=\alpha S, \quad
\bar {Q}=\sqrt\alpha{Q}, \quad
\bar  P=\frac{P}{\alpha}, \quad
\bar  V=\alpha^{3/2}V
\label{rescaling}
\end{equation}

\subsubsection{Criticality in grand canonical ensemble}

First, let us investigate the equation of state given by the isotherms $\bar P$ vs $\bar v$. We are going to use the following parametric expressions: $\bar P=\bar P(\bar \Lambda)$, $\bar T=\bar T(\bar \Lambda,\Phi,x_+)$, and the specific volume $\bar v=\bar v(\bar\Lambda,\Phi,x_+)$, which is
\begin{equation}
\label{vol_esp}
\bar v\equiv \frac{3}{2}\(\frac{\bar V}{\bar S}\)
=\frac{1}{\bar\eta}\(\frac{{x_+}+1}{x_+-1}\)
\end{equation}
We can solve the horizon equation to express $\bar\eta=\bar\eta(\bar\Lambda,\Phi,x_+)$  as
\begin{equation}
\bar\eta^2
=
\frac{3(x_{+}^2-1)-
(2\bar\Lambda+6\ln x_{+})x_{+}}
{{6(x_{+}-1)
\(2\Phi^2x_{+}-x_{+}+1\)}}
\label{free3}
\end{equation}
Notice that, by using the change of coordinates (\ref{change1}), we recover the specific volume for RN-AdS in the leading approximation: $v=2r_{+}+\mathcal{O}(r_+^{-1})$.

\begin{figure}[h]
\centering
\subfigure{\includegraphics[width=6.8 cm] {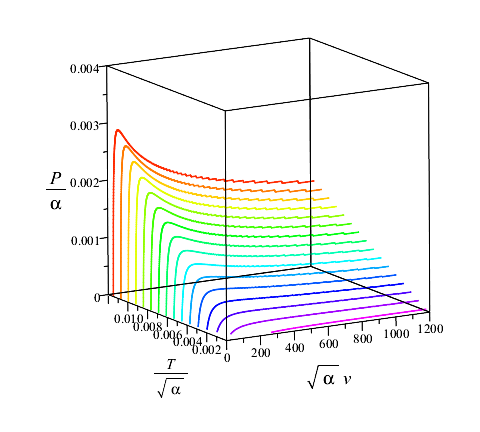}} \\
\subfigure{\includegraphics[width=6.8 cm] {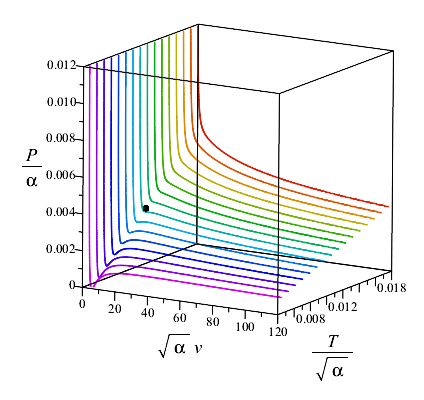}}\,
\subfigure{\includegraphics[width=7.4 cm] {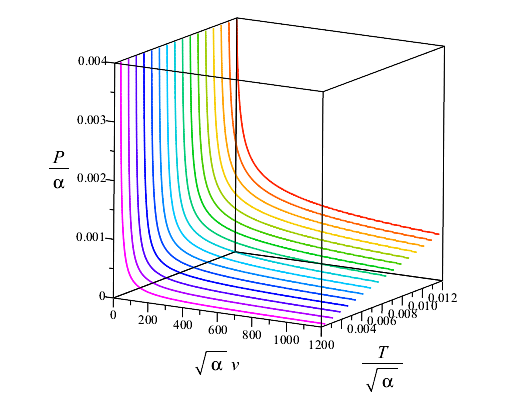}}\,
\caption{\small Equation of state represented for three distinct values of the gauge potential, namely $\Phi=0.6$, $\Phi=0.85$, and $\Phi=1.4$. Criticality occurs within  the interval $1/\sqrt2<\Phi<1$.}
\label{crgn2}
\end{figure}

\begin{figure}[h]
	\centering
	\subfigure{\includegraphics[width=14 cm]{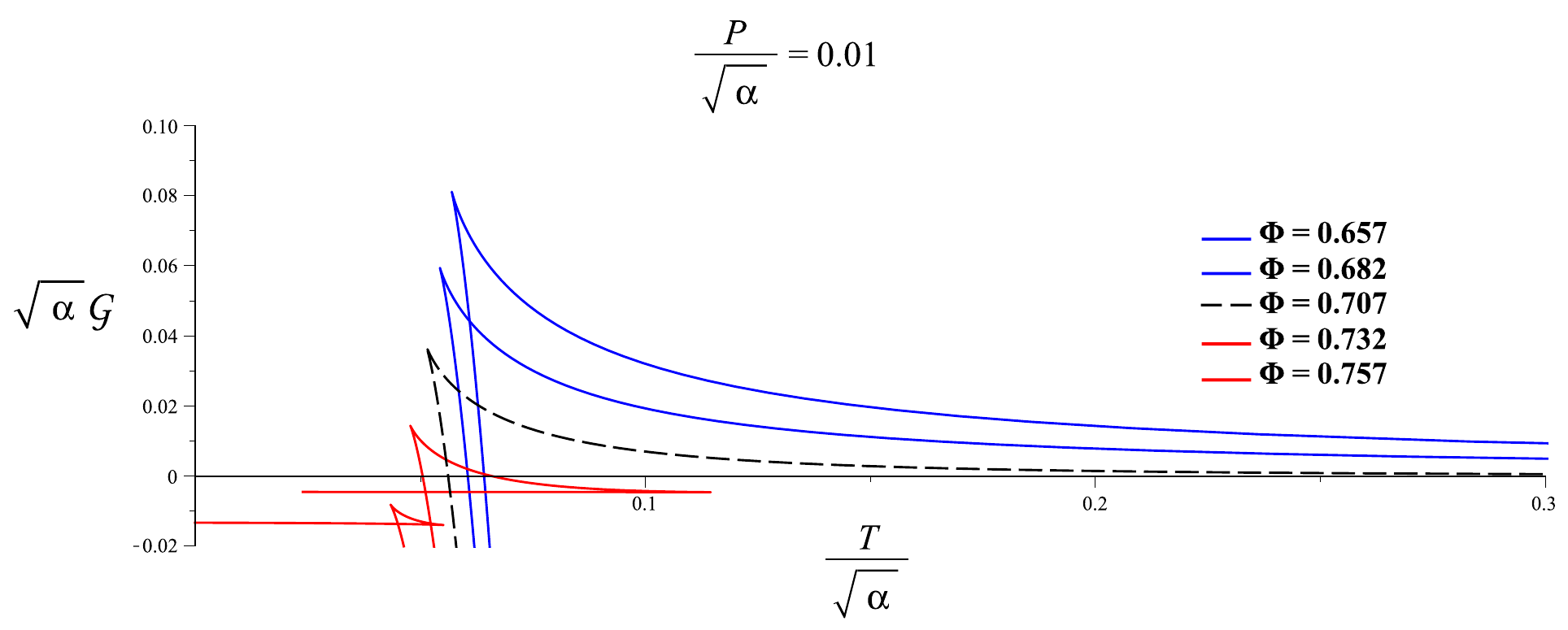}}
	\caption{\small Thermodynamic potential vs temperature, at fixed $\bar P=0.01$. The swallowtail develops when $\Phi$ increases, from below, to $1/\sqrt{2}$. The critical point (occurring on $\Phi\approx 0.778$) is not included in the figure.}
	\label{GTPPsi}
\end{figure}

\begin{figure}[h]
\centering
\subfigure{\includegraphics[width=5. cm] {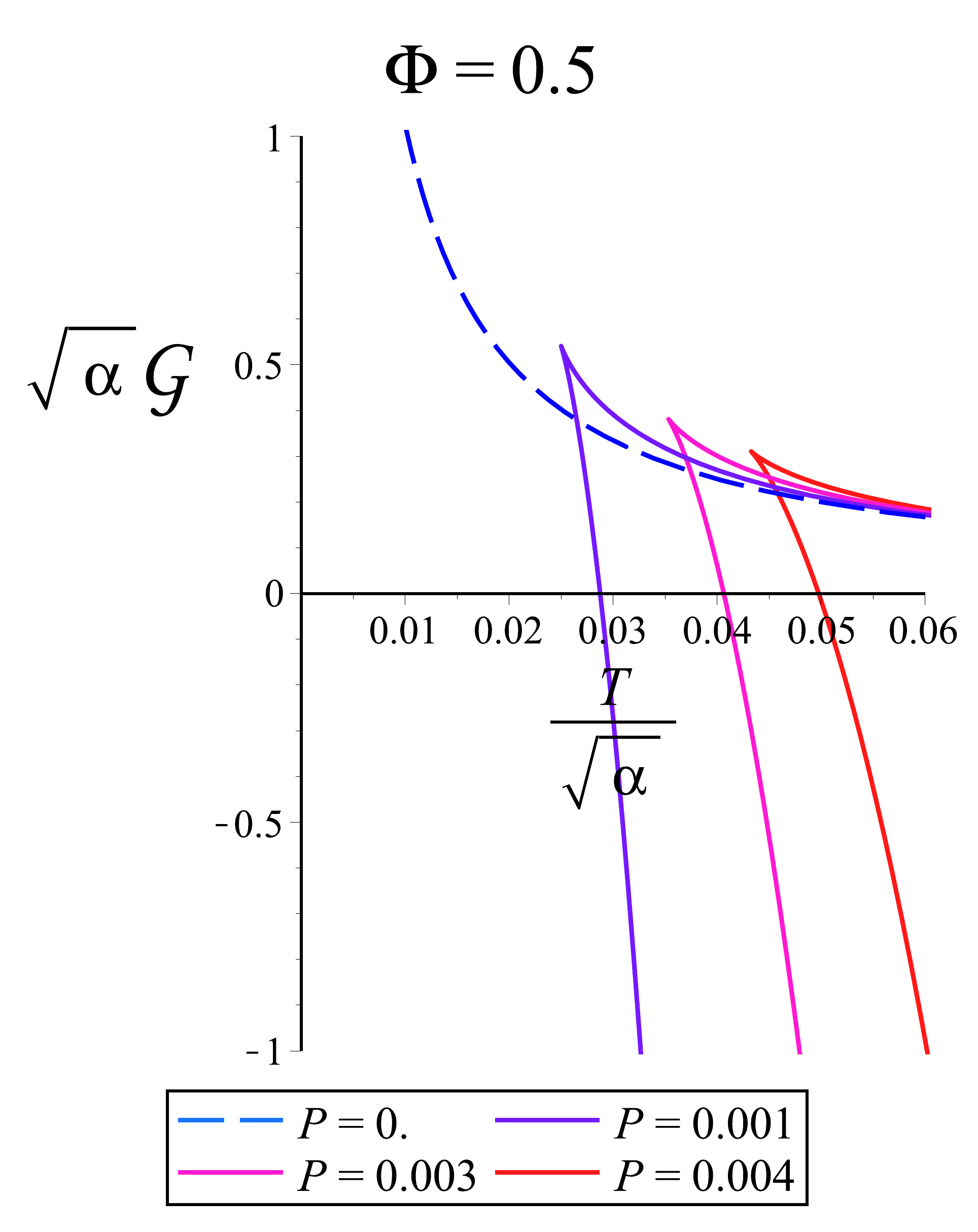}}	
\subfigure{\includegraphics[width=5. cm] {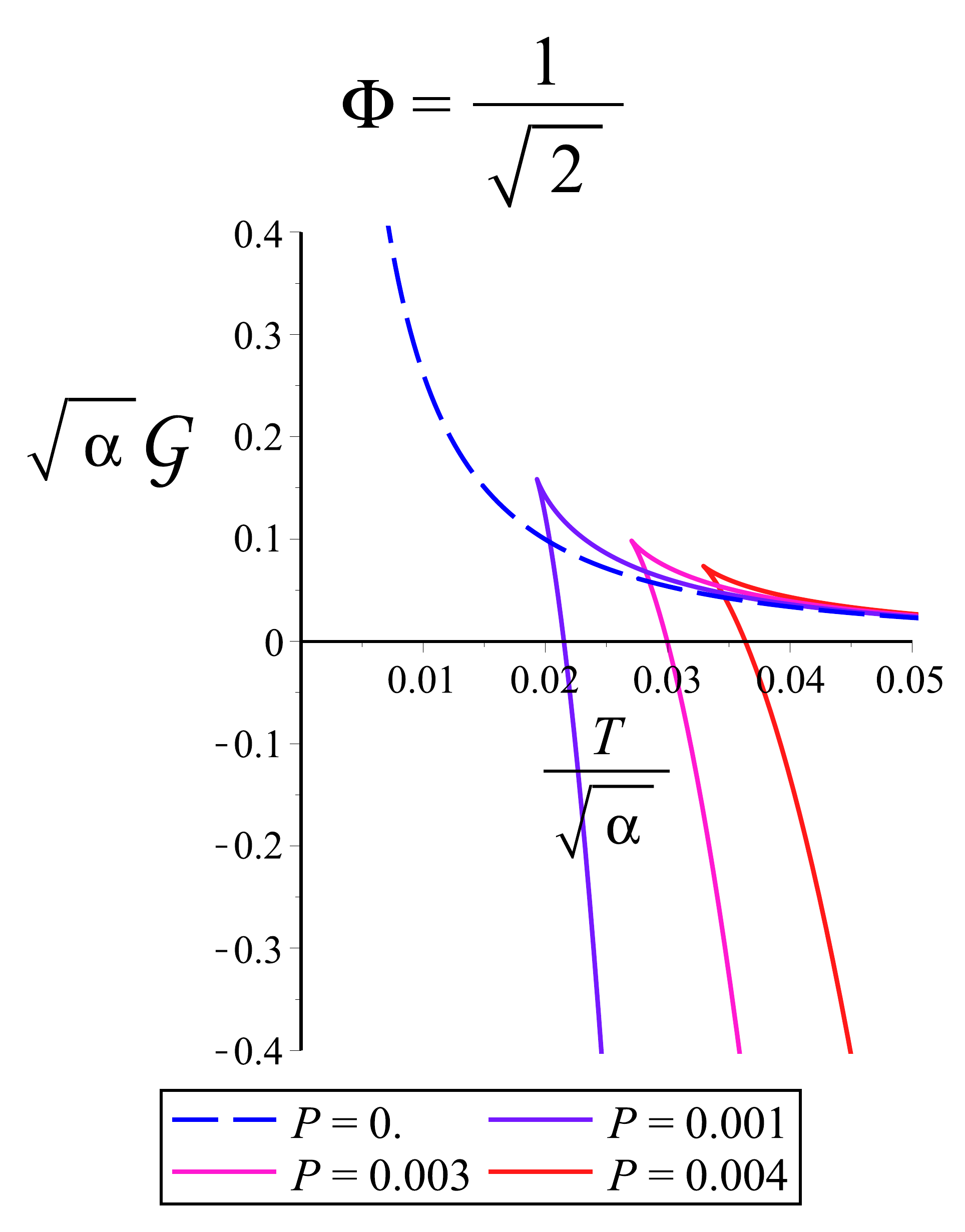}}
\subfigure{\includegraphics[width=5. cm] {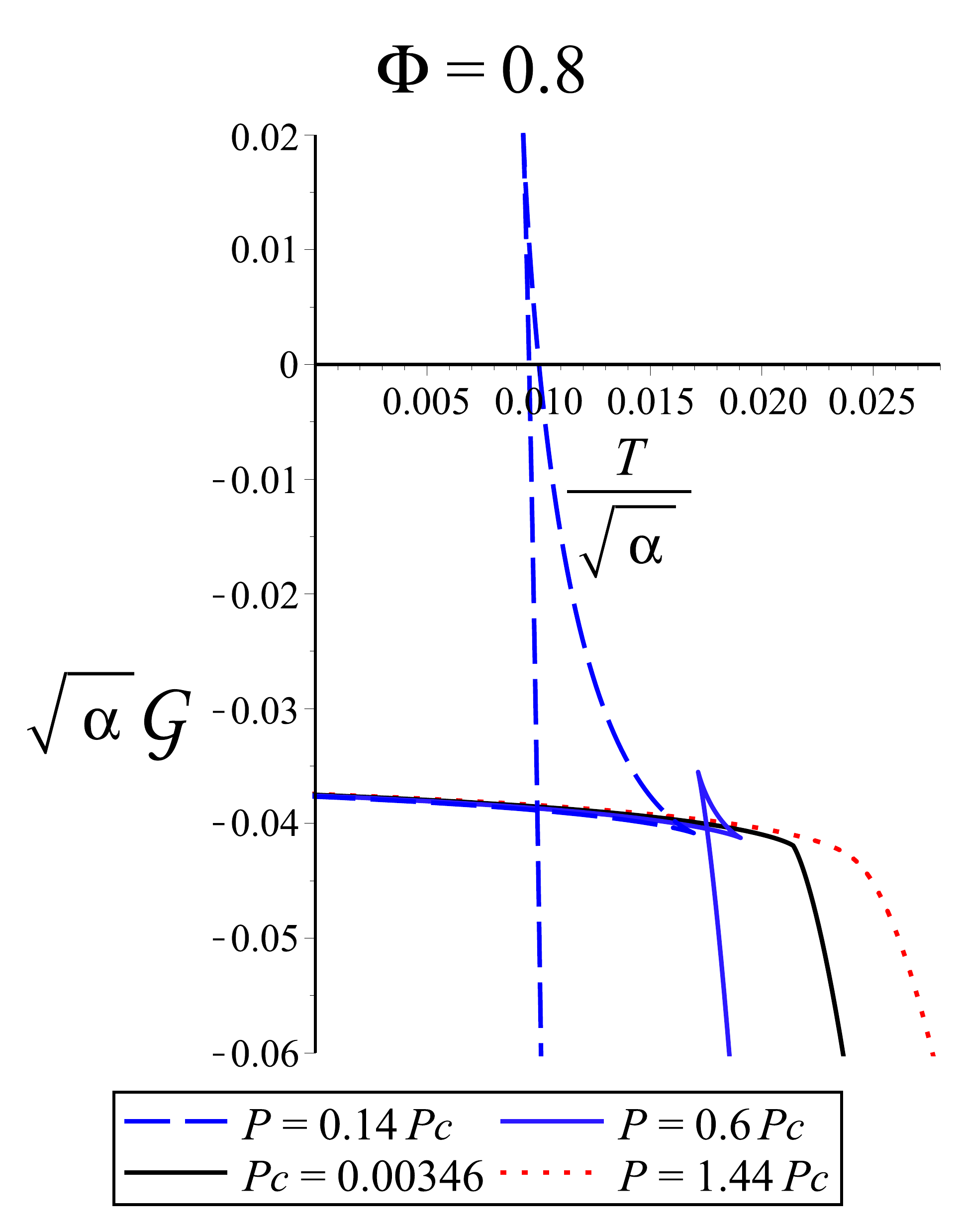}}
\caption{\small Thermodynamic potential in grand canonical ensemble, for three fixed $\Phi$. Given $\frac{1}{\sqrt{2}}<\Phi=0.8<1$, it is observed a critical isobar (black dash-dotted line) for which a critical point appears.}
\label{crgn3}
\end{figure}

\begin{figure}[h]
\centering
\subfigure{\includegraphics[width=6 cm] {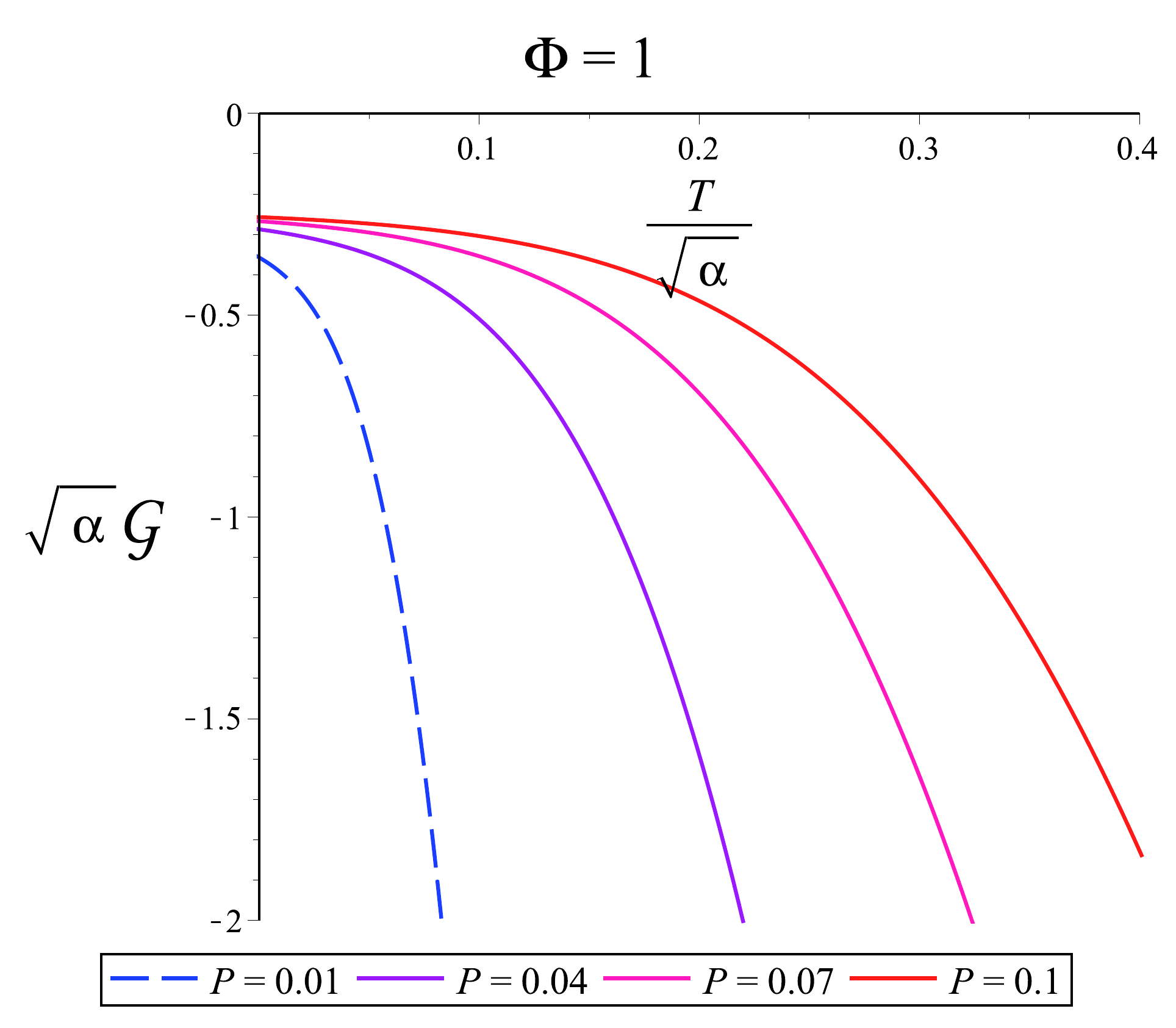}}\,\,\,\,\,\,\,\,\,\,\,\,\,\,\,\,\,\,\,\,\,\,\,\,\,\,\,\,\,\,
\subfigure{\includegraphics[width=6 cm] {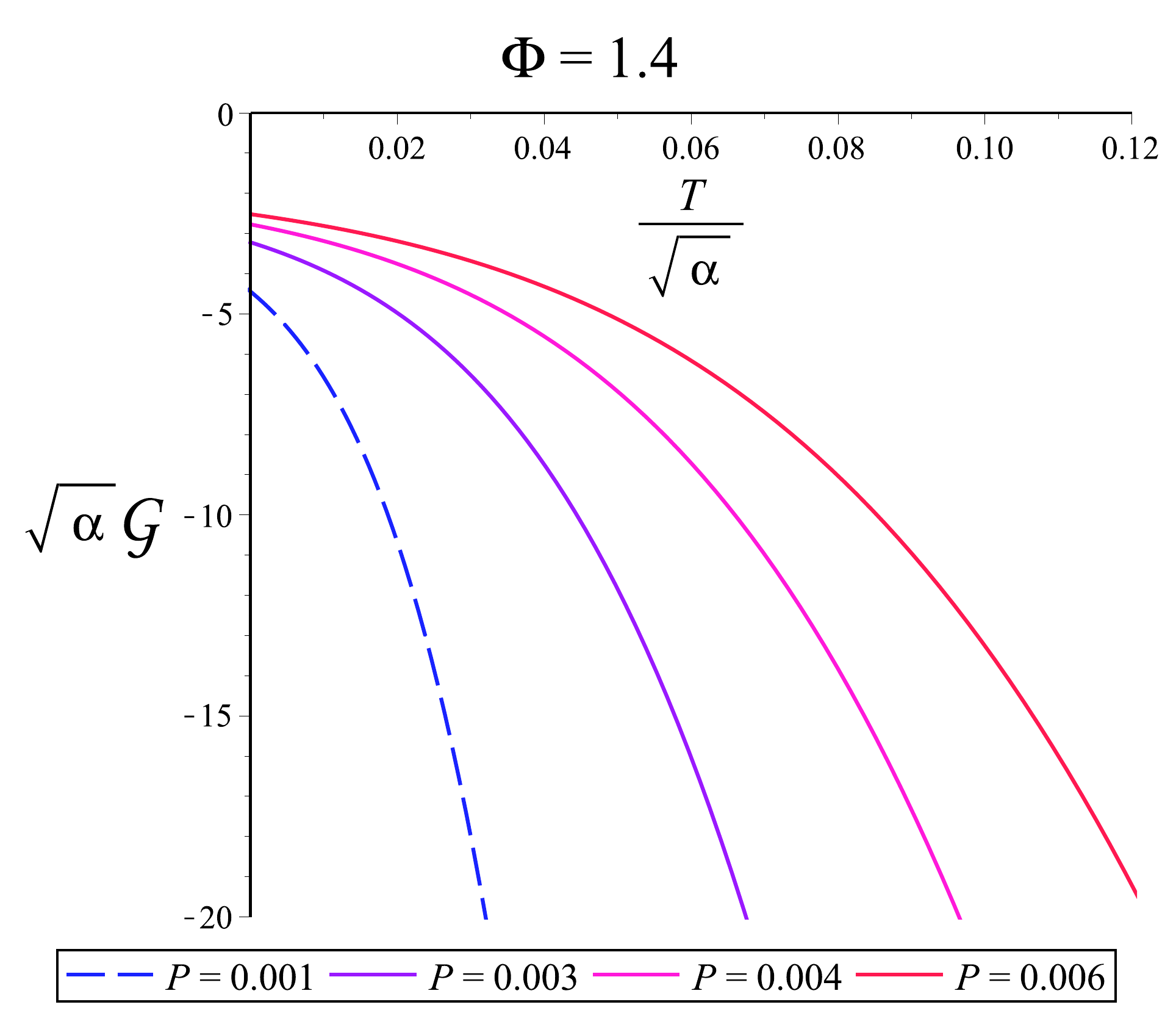}}
\caption{\small Thermodynamic potential in grand canonical ensemble, for $\Phi=1$ and $\Phi=1.4$.}
\label{crgn4}
\end{figure}

In Figure \ref{crgn2}, where we graphically represent the equation of state, one can  observe in one of the plots the existence of a critical point, characterized by the conditions $(\pa \bar P/\pa \bar v)_{\bar T,\Phi}=(\pa^2 \bar P/\pa \bar v^2)_{\bar T,\Phi}=0$.
The existence of an interval for the electric potential $\Phi$ for which there exists a non-trivial critical behaviour in grand canonical ensemble is, by its own, a novel property of these hairy charged black holes. At first sight, the nature of the critical point seems very similar with the one in RN-AdS, but in canonical ensemble (see Figure \ref{eqstateRN2}, in the previous section). The relevant quantities in terms of $(\bar P,\Phi,x_+)$ are
\begin{align}
\bar M&=
\frac{\Phi^2x_+^2}{(x_+-1)^2}
-\frac{1}{12\bar\eta^3},
\qquad
\bar T=
\frac{(x_{+}+2)\bar\eta\Phi^2}{2\pi}
-\frac{(x_+^2-1)\bar\eta}{4\pi x_+}
-\frac{(x_{+}-1)^2}{8\pi\bar\eta x_+}
\label{tempergc}
\end{align}
and the thermodynamic potential is obtained by computing the regularized on-shell action, as done previously, with the boundary condition $\left.\delta A_\mu\right|_{\pa\mathcal{M}}=0$. In terms of rescaled quantities, we  parametrically obtain
\begin{equation}
\mathcal{\bar G}(\bar P,\Phi,x_+)=
\frac{1}{24\bar\eta^3}
-\frac{2\Phi^2x_+^2-x_+^2+1}
{4(x_{+}-1)^2\bar\eta}
\label{gchairy}
\end{equation}
with $\bar\eta$ given by the equation (\ref{free3}) and $\bar\Lambda=-8\pi\bar{P}$.
This thermodynamic potential is depicted in Figure \ref{GTPPsi}, for fixed pressure and different values of $\Phi$, and in Figure \ref{crgn3} and \ref{crgn4} for fixed $\Phi$ and different values of $P$. A careful analysis of the region of criticality provides the following range of validity for the conjugate potential: 
\begin{equation}
\frac{1}{\sqrt2}<\Phi<1
\label{windows}
\end{equation}


In Figure \ref{GTPPsi}, the new feature of `snapping swallowtails'  appears when the value of the conjugate potential  crosses the critical value $\Phi_c=1/\sqrt{2}$, after which it resembles the behaviour of the RN-AdS black hole in the canonical ensemble where there exists a critical charge. We can observe that the thermodynamic potential (\ref{gchairy}) simplifies drastically for this critical value, $\Phi_c=1/\sqrt{2}$. For the interval $0<\Phi\leq 1/\sqrt{2}$ there is a Hawking-Page phase transition at $\mathcal{G}=0$, between the large black hole and pure thermal radiation. A simple argument to check that, indeed, the thermodynamic potential $\mathcal{G}$ can vanish is that, from equation (\ref{free3}), we observe that $\eta$ diverges when the factor $2\Phi^2x_{+}-x_{+}-1$ goes to zero. This limit occurs for $x_+=-\frac{1}{2}
(\frac{1}{\Phi^2-\frac{1}{2}})$, that is, for $\Phi\leq\frac{1}{\sqrt{2}}$ (in positive branch). A drastic change appears when $\Phi>\frac{1}{\sqrt{2}}$, namely a swallowtail can develop as can be seen in Figure \ref{GTPPsi}. At the swallowtail intersection there is a large-small black hole phase transition. The particular value $\Phi=1$, which is also relevant for our analysis,  is more easily understood because, as can be observed in Figure \ref{crgn4}, the thermodynamic potential cannot vanish. Indeed, if we compute $\bar{\eta}$ from equation (\ref{gchairy}) when $\mathcal{\bar G}=0$ and $\Phi=1$ and replace it in equation (\ref{tempergc}), we obtain a negative temperature, a situation that is physically impossible. If we keep the chemical potential fixed in the interval $\frac{1}{\sqrt2}<\Phi<1$ and vary the pressure, then there is a critical pressure  for the  snapping of the swallowtail similar to that of the snapping swallowtail in accelerating black hole thermodynamics \cite{Abbasvandi:2018vsh}.

\begin{figure}[h]
\centering
\subfigure{\includegraphics[width=6. cm] {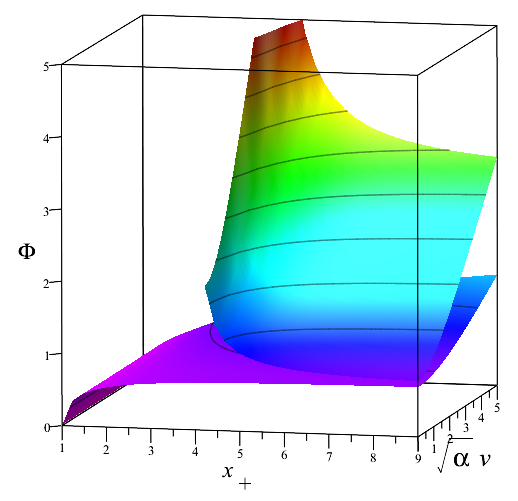}}	\,\,\,\,\,\,\,
\subfigure{\includegraphics[width=6.1 cm] {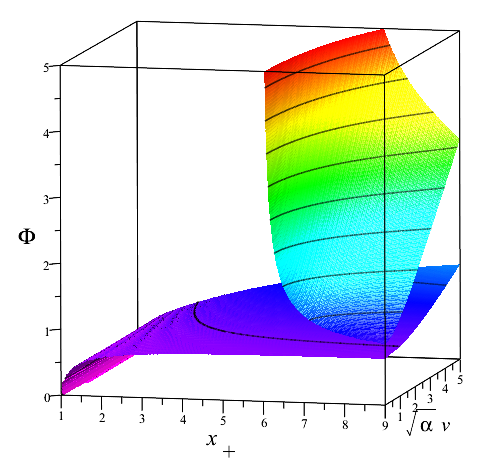}}
\caption{\small At the left hand sid the surface where $\(\pa \bar{P}/\pa\bar{v}\)_{\Phi,\bar{T}}=0$    is shown. At the right hand side we show the surface where $\(\pa^2 \bar{P}/\pa\bar{v}^2\)_{\Phi,\bar{T}}=0$.}
\label{crit1}
\end{figure}

\begin{figure}[h]
\centering
\subfigure{\includegraphics[width=6. cm] {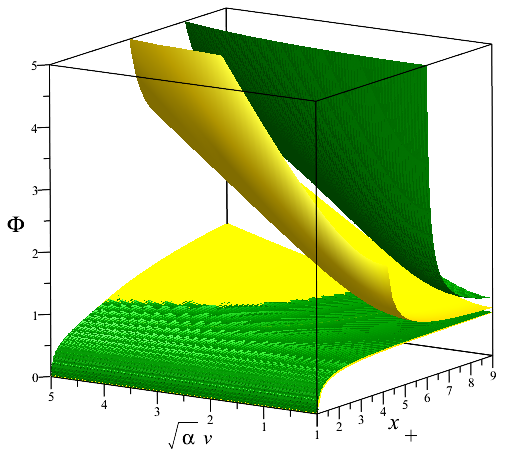}}\,\,\,\,\,\,\,\,\,\,\,\,\,\,
\subfigure{\includegraphics[width=6. cm] {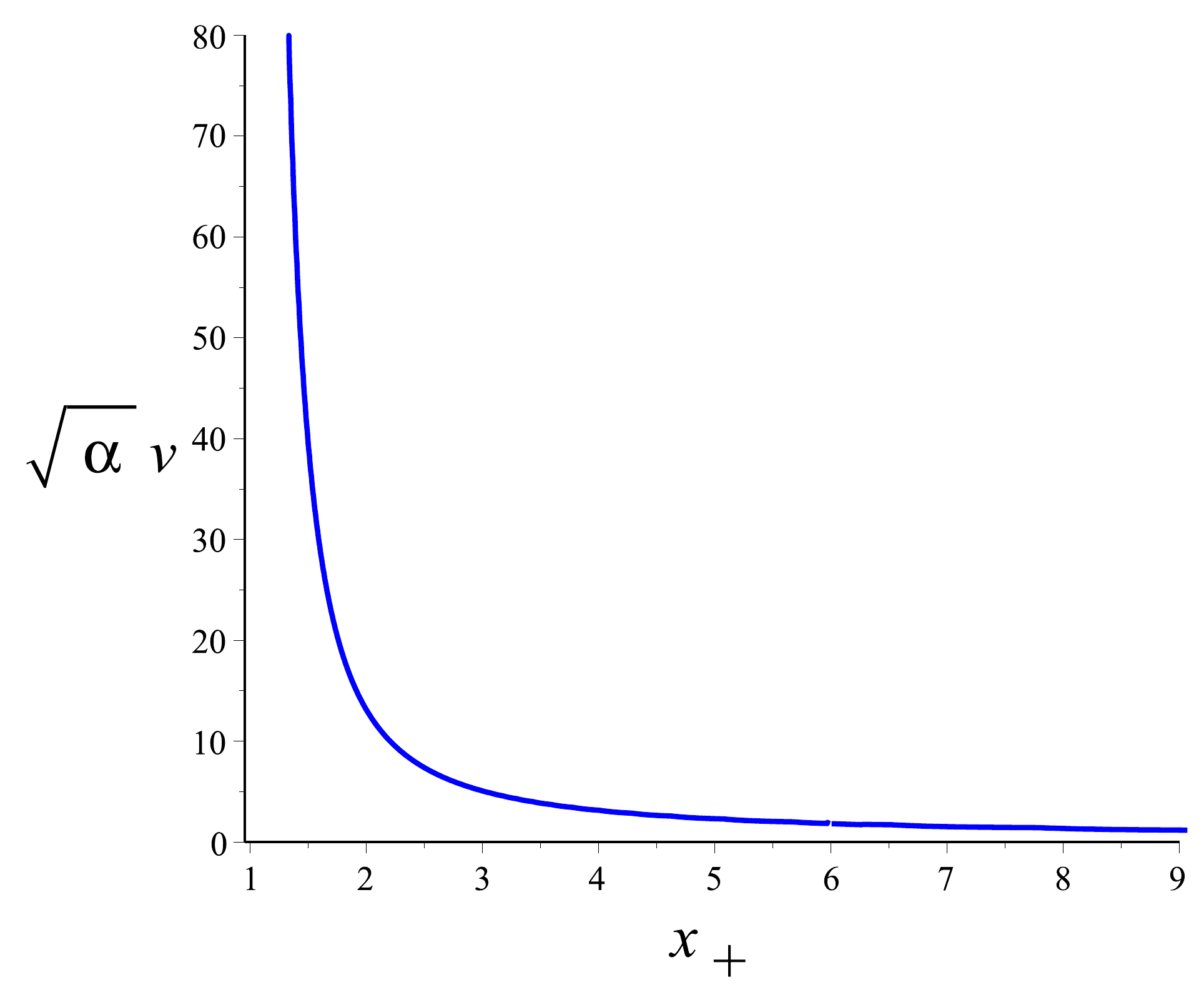}}
\caption{\small The left hand side depicts simultaneously the two surfaces corresponding to the conditions for criticality (Figure \ref{crit1}); the surface in yellow corresponds to $\(\pa \bar{P}/\pa\bar{v}\)_{\Phi,\bar{T}}=0$ and the surface in green to $\(\pa \bar{P}^2/\pa\bar{v}^2\)_{\Phi,\bar{T}}=0$.
The intersection occurs inside the interval $1/\sqrt{2}<\Phi<1$, which is shown in the right hand side, in the ($\bar v$,$x_+$)  plane. }
\label{line}
\end{figure}

Let us present further arguments that support our findings. From the equations (\ref{vol_esp}) and (\ref{free3}), we can obtain $\bar P=\bar P(\Phi,\bar v,x_+)$. The derivative $\(\pa \bar P/\pa \bar v\)_{\Phi,\bar T}$ is given by
\begin{equation}
\(\frac{\pa \bar P}{\pa \bar v}\)_{\Phi,\bar T}=
\(\frac{\pa \bar P}{\pa \bar v}\)_{\Phi,x_+}
+\(\frac{\pa \bar P}{\pa x_+}\)_{\Phi,\bar v}
\(\frac{\pa x_+}{\pa \bar v}\)_{\Phi,\bar T}
\end{equation}
where $\(\pa x_+/\pa \bar v\)_{\Phi,\bar T}$ is obtained from equation (\ref{temperh}). To get the second derivative $\(\pa^2 \bar P/\pa \bar v^2\)_{\Phi,\bar T}$, let us define $\bar P_2\equiv \(\pa \bar P/\pa \bar v\)_{\Phi,\bar T}$ and then we obtain
\begin{equation}
\(\frac{\pa^2 \bar P}{\pa \bar v^2}\)_{\Phi,\bar T}=
\(\frac{\pa \bar P_2}{\pa \bar v}\)_{\Phi,x_+}
+\(\frac{\pa \bar P_2}{\pa x_+}\)_{\Phi,\bar v}
\(\frac{\pa x_+}{\pa \bar v}\)_{\Phi,\bar T}
\end{equation}
It is possible to explicitly get, though cumbersome to write down, parametric expressions for the conditions of criticality, $\(\pa\bar P/\pa \bar v\)_{\Phi,\bar T}=0$ and $\(\pa^2 \bar P/\pa \bar v^2\)_{\Phi,\bar T}=0$. Let us analyze the first condition for the first derivative equal to zero, which is
\begin{equation}
{\Phi}^{4}
+g_1(\bar{v},x_+)\Phi^2+g_2(\bar{v},x_+)=0
\label{ct1}
\end{equation}
with
\begin{align}
g_1(\bar v,x_+)&=-\frac{(x_{+}-1)
\[\(x_{+}-1\)^3\bar{v}^2
+2\(x_{+}-3\)\(x_{+}+1\)^2\]}
{2x_+\,\(x_{+}-4\)\(x_{+}+1\)^2}\\
g_2(\bar{v},x_+)&=\frac{\(x_{+}-1\)^4
\[(x_{+}-1)^4v^4+4\(x_{+}-1\)\(x_{+}+1\)^3v^2
+4\(x_{+}+1\)^4\]
}{16x_+^3\(x_{+}-4\)\(x_{+}+1\)^4}
\end{align}
The second condition, for the second derivative, is more complicated, but schematichally it looks like $\Phi^8+h_1\Phi^6+h_2\Phi^4+h_3\Phi^2+h_4=0$, for $h_{i}=h_i(\bar v,x_+)$. 
In Figure \ref{crit1}, we illustrate separately the surfaces of points where the first and second derivatives vanish. Despite both having a similar shape, they do not meet for the whole range of $\Phi$. In Figure \ref{line} we  depict, in the ($\bar v$,$x_+$)  plane  the curve along which both conditions for criticality hold.

\begin{figure}[h]
	\centering
	\subfigure{\includegraphics[width=7 cm]
		{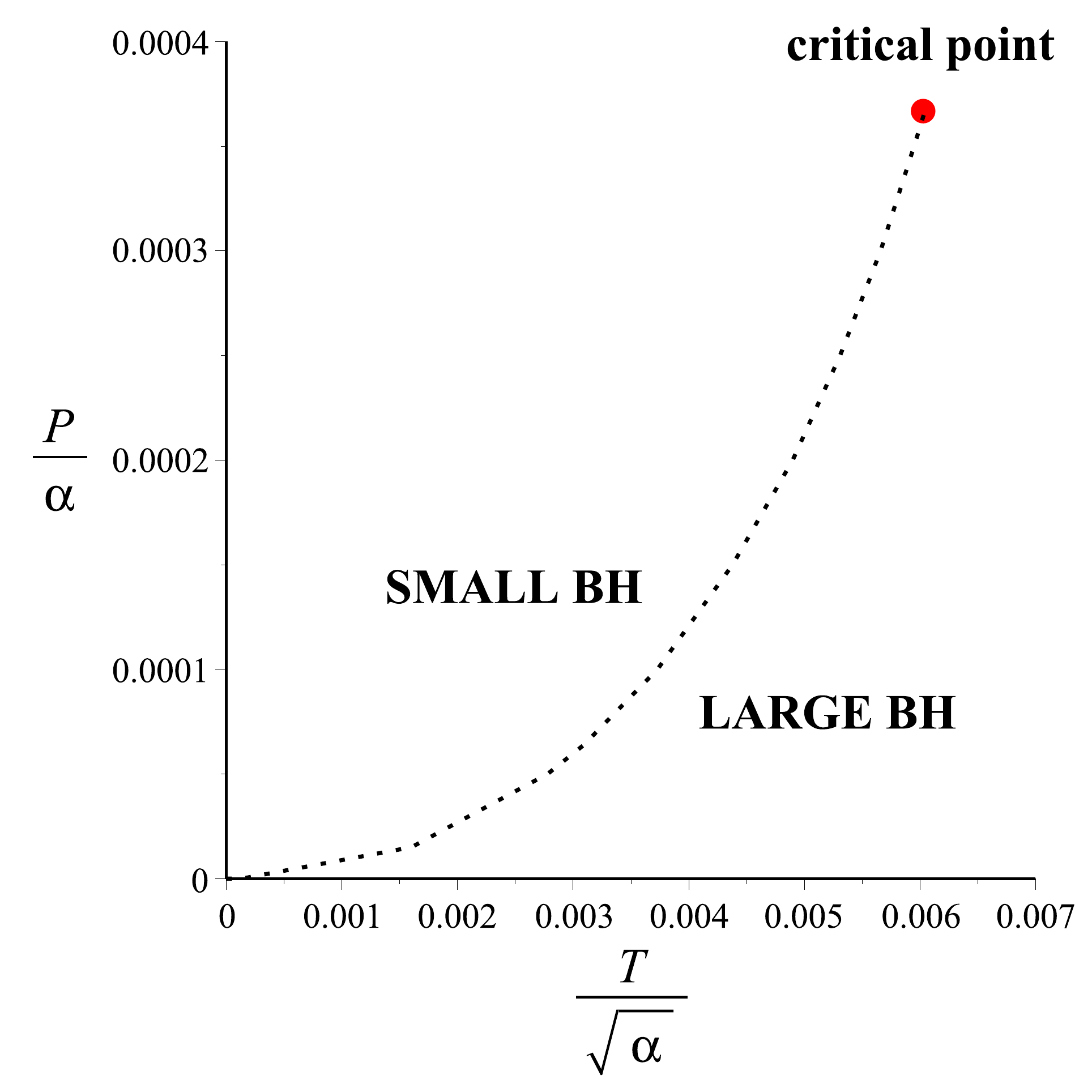}}	\,\,\,\,\,\,\,\,\,
	\subfigure{\includegraphics[width=7 cm]
		{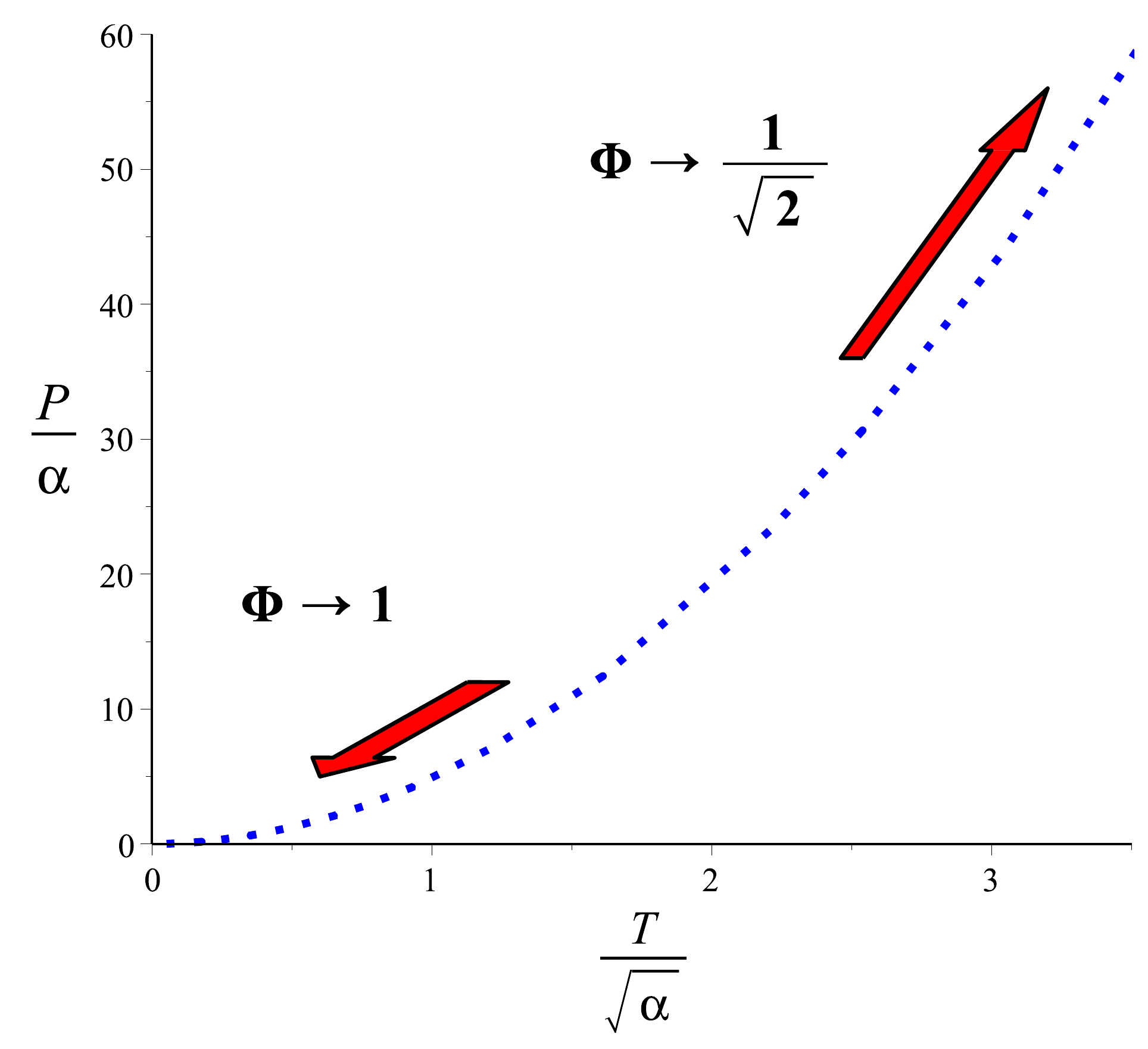}}	
	\caption{\small Left hand side: The dotted line is the coexistence line, finishing at the critical point (the red point), in the grand canonical ensemble for a fixed $\Phi=0.85$. Right hand side: trajectory of the critical points with respect to $\Phi$. In the limit $\Phi=1$, the critical point goes to $(0,0)$. In the limit $\Phi=1/\sqrt2$, the critical point correspond to large values of $\bar P$ and $\bar T$.}
	\label{coexGC}
\end{figure}

Also, in Figure \ref{coexGC}, on the plot in  the left hand side, we show the coexistence line for the particular value $\Phi=0.85$. This line is similar to the coexistence line for RN-AdS in the canonical ensemble. On the same figure, but in the plot from the right hand side,  the trajectory of the critical points for the range of criticality $1/\sqrt{2}<\Phi<1$ is shown.

\subsubsection{Criticality in canonical ensemble}
The canonical ensemble corresponds to a fixed charge and it can be obtained from a Legendre transform of the thermodynamic potential in grand canonical ensemble. The black holes are thermodynamic systems, but also geometric objects, and from a geometric point of view the Legendre transform between ensembles is equivalent to adding the boundary term (\ref{boundary2}) that will contribute non-trivially to the Euclidean action (for the corresponding boundary conditions of the canonical ensemble). Since fixing $Q$ is equivalent to fixing $\bar{Q}$, provided $\alpha$ is fixed, it is convenient to express the thermodynamic quantities in terms of $\bar{Q}$, instead of $Q$. Using  (\ref{rescaling}) we obtain
\begin{align}
\bar M&=\frac{12\bar{Q}^2\bar\eta^4-1}{12\bar\eta^3} ,\quad
\bar T=\frac{(x_{+}-1)^2}{8\pi \bar\eta x_+}
\[-1-2\bar\eta^2\(\frac{x_{+}+1}{x_{+}-1}\)
+4\bar\eta^4\bar{Q}^2\(\frac{x_{+}+2}{x_+}\)\] \\
\bar S&=\frac{\pi x_+}{\bar\eta^2\(x_{+}-1\)^2} ,\quad
\Phi=\bar Q\bar\eta\(\frac{x_{+}-1}{x_+}\) ,\quad
\bar{V}=\frac{2\pi x_+}{3\bar\eta^3}
\frac{x_{+}+1}{(x_{+}-1)^3}
\end{align}
where $\bar{\eta}=\bar{\eta}\(\bar{Q},\bar{P},x_+\)$ can be obtained from the horizon equation \eqref{horizon}
\begin{equation}
\bar\eta=\frac{1}{2\bar{Q}}\sqrt{\frac{x_+}{x_{+}-1}}
\[1+\sqrt{1+\frac{4\(16\pi x_+\bar{P}-6x_+\ln{x_+}+3x_+^2-3\)\bar{Q}^2}
{3x_+\(x_{+}-1\)}}\;\]^\frac{1}{2}
\end{equation}
\begin{figure}[h]
	\centering
	\subfigure{\includegraphics[width=6.6 cm]{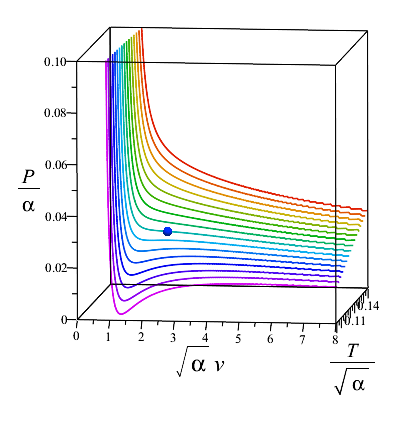}}\,\,\,\,\,\,\,
	\subfigure{\includegraphics[width=7.05 cm]{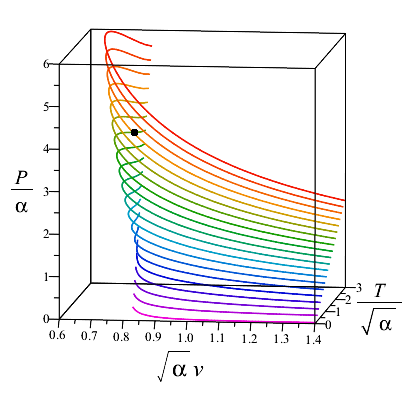}}
	\caption{\small Equation of state, given $\bar Q=0.67$. The two plots correspond to different regions of the same phase space. \textbf{Left}: In this region, we observe a  critical point similar to that in RN-AdS. \textbf{Right:} Here, we observe a \textit{purely hairy} critical point. Note that, for a given temperature, as the pressure increases, the specific volume decreases until it reaches a minimum and the surface develops a bend. Next both $P$ and $v$ increase until, for a critical isotherm,  the hairy critical point emerges.   
}
	\label{pv1}
\end{figure}
We have used (\ref{vol_esp}) to eliminate $\bar{\eta}$ in order to express $\bar{T}=\bar{T}(\bar{Q},\bar{P},x_+)$ and $\bar{v}=\bar{v}(\bar{Q},\bar{P},x_+)$. Since it is not possible to get an analytic expression $\bar T=\bar T(\bar P,\bar v)$, we plot the equation of state in Figure \ref{pv1} for a particular value of electric charge, $\bar{Q}=0.67$. 

\begin{figure}[h]
	\centering
	\subfigure{\includegraphics[width=6.8 cm]{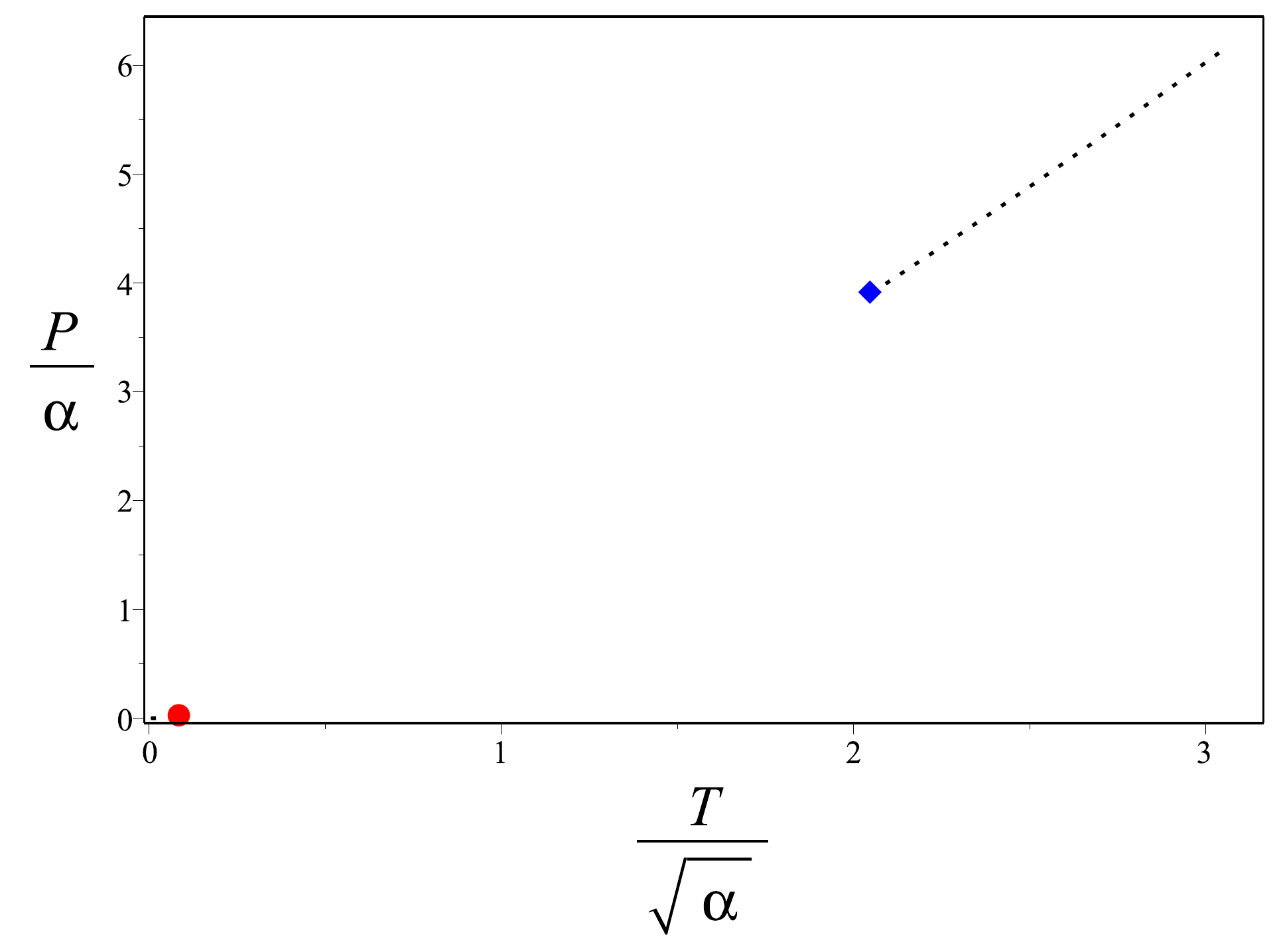}} \,\,\,\,\,\,\,
	\subfigure{\includegraphics[width=6.8 cm]{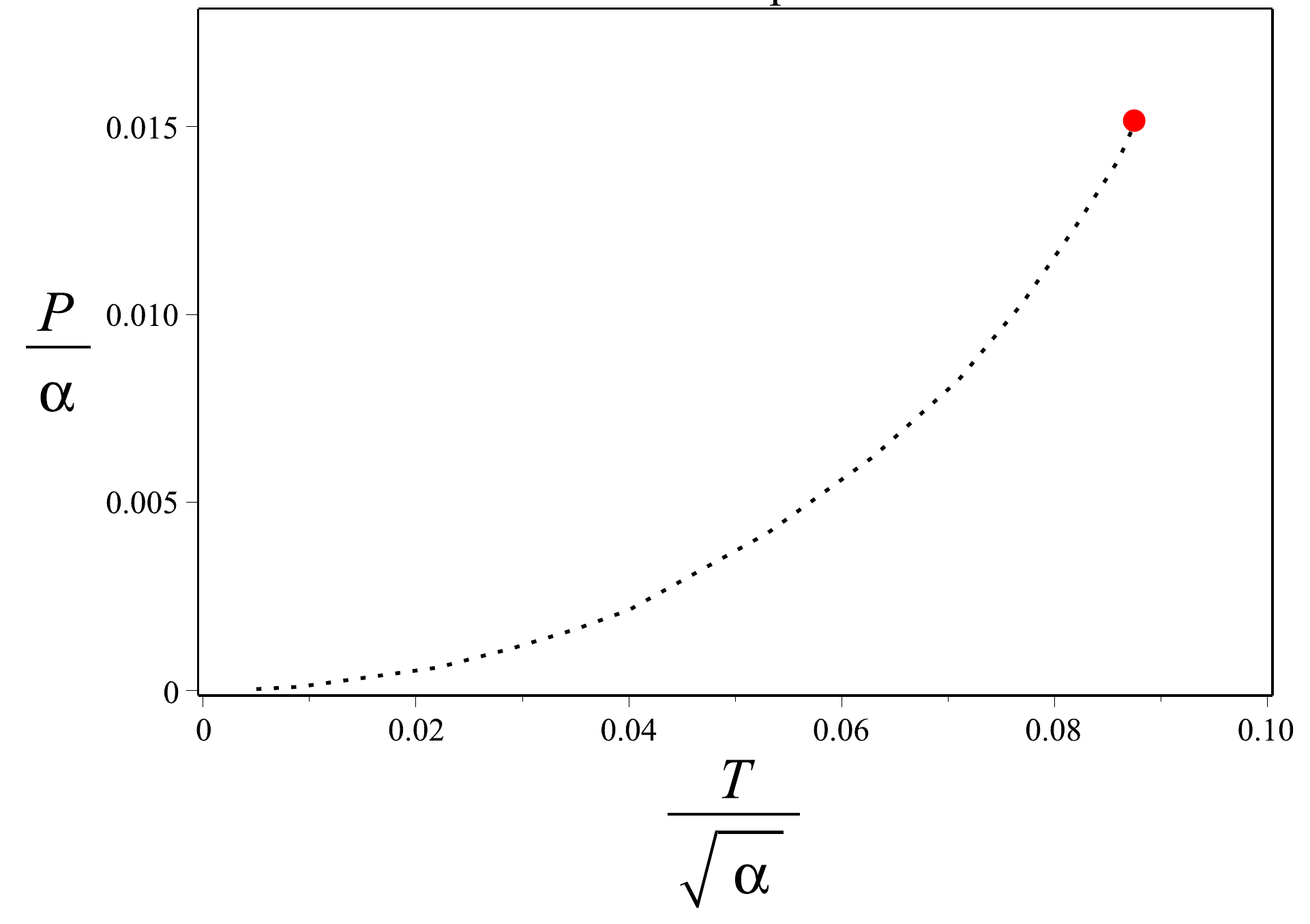}}
	\caption{\small \textbf{Left}: It was fixed $\bar{Q}=0.67$. The red point, at the left, is the RN-AdS-like critical point. The second blue point, at the right, is the hairy critical point. The dotted lines indicate the coexistence line. \textbf{Right}: Magnification near the RN-AdS-like critical point and its coexistence line.
	}
	\label{PTcurves}
\end{figure}

\begin{figure}[h]
	\centering
	\subfigure{\includegraphics[width=12 cm]{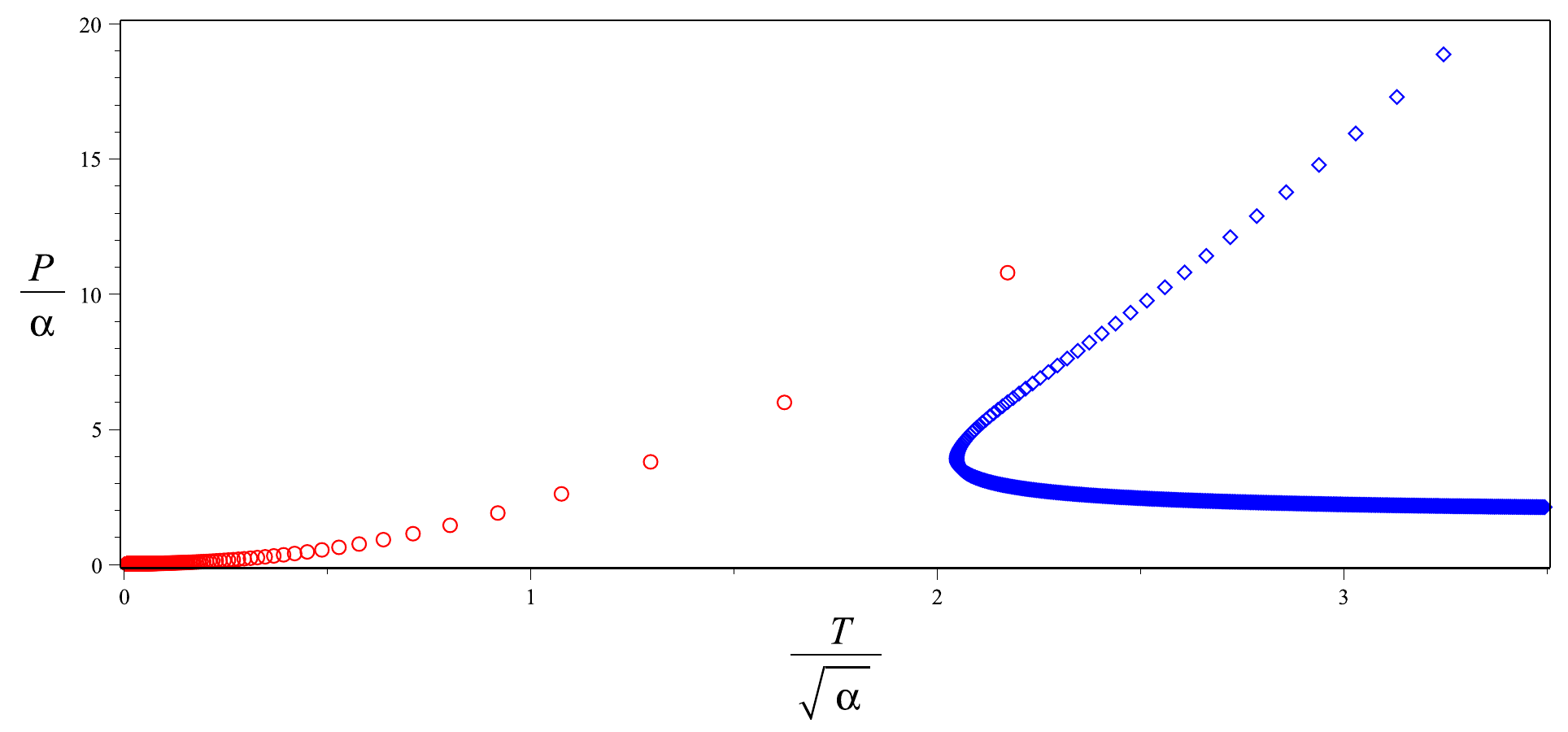}}
	\caption{\small Trajectory of both critical points, parametrized by $\bar{Q}$. Red points, at the left, correspond to RN-AdS-like critical point and blue ones, at the right, to the hairy critical points. The zone where the points are more concentrated corresponds to the limit $\bar Q\rightarrow\infty$, while the more diluted zone corresponds to the limit $\bar Q\rightarrow 0$.}
	\label{PTtray}
\end{figure}

One interesting novel aspect of the equation of state in the canonical ensemble is the existence of two configurations satisfying simultaneously the criticality conditions 
\begin{equation}
\(\frac{\pa\bar{P}}{\pa\bar{v}}\)
_{\bar{T},\bar{Q}}=0,\qquad
\(\frac{\pa^2\bar{P}}{\pa\bar{v}^2}\)
_{\bar{T},\bar{Q}}=0
\end{equation}
These two critical points simultaneously exist for any value of $\bar Q$. One of these points is the expected RN-AdS-like critical point at large volume and low temperature; the other one is  at small volume and high temperature; it is a new feature of this solution and presents different properties. In Figure \ref{PTcurves}, we plot the coexistence curves for both phases for a given value of $\bar Q$. The trajectories of the critical points in $P$--$T$ space as a function of charge are depicted in Figure \ref{PTtray}. 

\begin{figure}[h]
	\centering
	\subfigure{\includegraphics[width=13 cm]{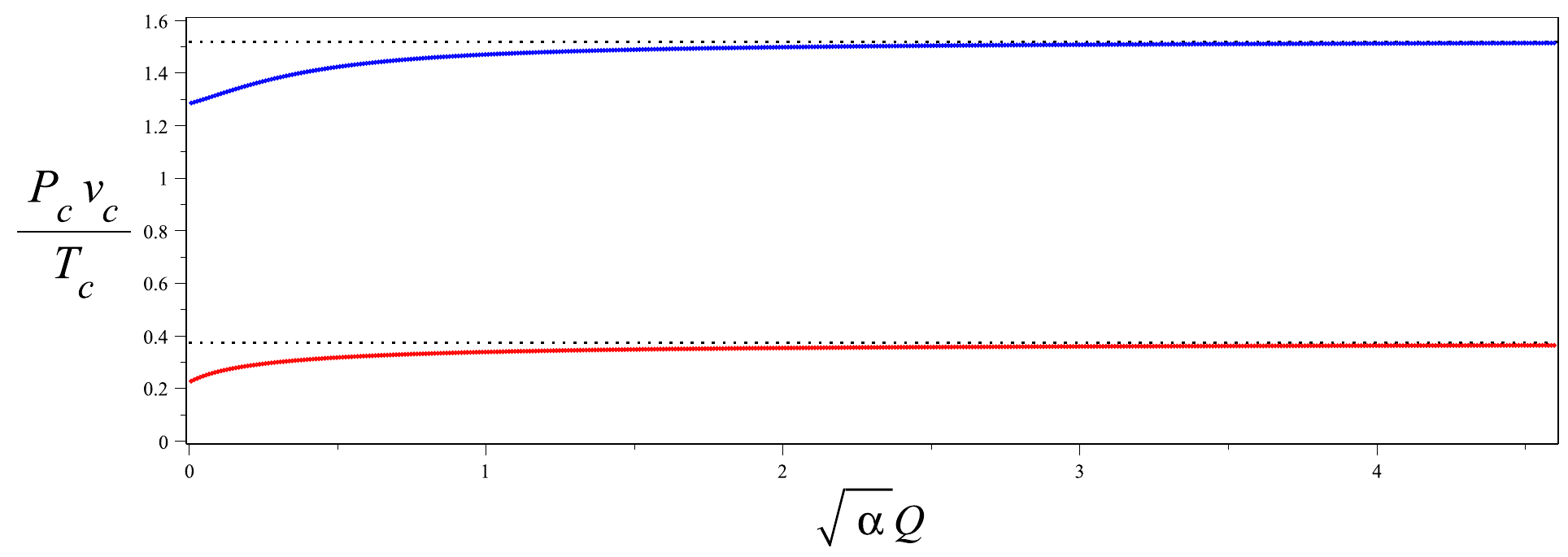}}
	\caption{\small Critical compresibility factor $Z_c\equiv P_cv_c/T_c$ as function of $\bar{Q}$. The red curve below represents the RN-AdS-like critical points, and the blue one above, the hairy critical points. As expected, the RN-AdS-like critical approaches to $Z_c={3}/{8}$ when $\sqrt{\alpha}Q\gg 1$, while the hairy critical points  approaches to $Z_c\approx 1.52$.}
\label{compress}	
\end{figure}
From Figure \ref{PTtray}, we observe that, for increasing charge (starting from $\bar Q=0$, corresponding to the more diluted zone), the new critical point --- the curve in blue, at the right --- decreases in pressure and temperature up to some
minimum temperature, when the point $(P_c,T_{c-min})$ is reached. As charge further increases (towards the more concentrated zone), the critical point moves to increasingly high temperature but still decreasing pressure.

The RN-AdS-like critical compressibility factor approaches   $Z_c=3/8=0.375$ in the limit $Q\gg \alpha^{-1/2}$, and to $Z_c\approx 0.22$ when $\bar Q=0$, as shown in Figure \ref{compress}.
It seems a mere coincidence that several critical compressibility factors for different chemical substances are located between these two numerical values (see, e.g., \cite{Kulinskii2013,Wei2013}), but the fact that $Z_c$ depends on the charge suggests that $\bar Q$ plays the role of a \textit{fluid parameter}, i.e., its value modifies the nature of the fluid in the dual theory. This observation could be important in the context of AdS/CFT duality. 

\section{Discussion}
\label{Discuss}
Scalar fields arise naturally as moduli in the context of string theory when considering specific compactifications and the low energy limit. The dilaton can couple non-trivially to the electromagnetic field   \cite{Garfinkle:1990qj}. Since string theory is a fundamental theory that could describe   nature, it is important to understand the influence of the dilaton on the thermodynamic behaviour of black holes. Furthermore, in the context of gauge-gravity duality, the study of asymptotically Anti-de Sitter black holes is important because it may provide specific insights into new phenomena and phases of matter of the dual field theories. 

For these reasons we present  a possible new interpretation of the `extended thermodynamics' program within string theory.
 The traditional interpretation within general relativity is that, once the cosmological constant becomes dynamical, the first law should be modified by a new term, $PdV$, and the mass of the black hole becomes, in fact, the enthalpy of spacetime \cite{Kastor:2009wy}. Clearly, this interpretation calls for a more careful scrutiny in the context of gauge-gravity dualities.\footnote{Despite some results \cite{Karch:2015rpa,Sinamuli:2017rhp},  the interpretation in AdS-CFT duality is still unsettled.} However, even in the traditional sense, there is some tension in modifying the first law of thermodynamics when a parameter of the theory is varying. The first law is a statement between two configurations/solutions in thermodynamic equilibrium. Modifying the cosmological constant is like modifying the theory and the correct picture is rather that a single configuration (the black hole) should be considered  in different theories characterized by different cosmological constants. 

In clarifying this puzzle, we would like to briefly present the recent resolution of a similar puzzle for the scalar charges \cite{Astefanesei:2018vga}. Since in string theory the dilaton is a modulus whose expectation value controls the (dimensionless) string  constant, $g_s$, it makes sense to understand how a variation of its asymptotic value affects the thermodynamics of asymptotically flat hairy black holes. Such a study was performed in \cite{Gibbons:1996af} with a surprising result. That is, the first law of hairy black hole thermodynamics should be supplemented with contributions from the scalar fields. However, recently, it was shown in  \cite{Astefanesei:2018vga} that when the correct variational principle is considered, the total (quasilocal) energy has a new contribution that depends of the asymptotic value of the scalar field and there is no need  modify the first law by adding new terms depending on the non-conserved scalar charges. This fits well with the string theory interpretation that changing the boundary condition for the dilaton is equivalent with changing the string coupling.

A similar reasoning can be made when the cosmological constant becomes dynamical. Once embedded in string/M theory \cite{Cvetic:1999xp}, the radius of AdS is related to the radius of the external sphere. Therefore, changing the cosmological constant is equivalent with a geometrical process, namely one with a (dynamical) variation of the volume (modulus) of the five/four/seven-sphere depending if one considers $AdS_5$, $AdS_7$ or $AdS_4$. So, this hints to a new interpretation for which the same black hole configuration (characterized by its conserved charges) in AdS can be embedded in different theories obtained by various compactifications. Therefore, the total energy of the black hole computed in a different theory should be, obviously, different. Then, once embedded in string theory, the criticality phenomenon comes with a critical pressure that in fact corresponds geometrically to a specific `critical compactification'. An important observation is that the potential we have considered (\ref{dilaton}) comes with another independent parameter, $\alpha$. We can consider also $\alpha$ as a dynamical parameter and, in a particular case, we can exactly cancel out the $PdV$ term by a specific variation of $\alpha$. Then, despite the fact that the cosmological constant is not kept constant, there is no extra contribution in the first law as it should if the interpretation of  \cite{Kastor:2009wy} would be correct.

There is an important difference with the proposal of \cite{Kastor:2009wy}, which is in the context of general relativity where the boundary conditions are in general kept fixed. In the first law only variations of the conserved charges (integration constants) of a solution enter, not variations of the parameters of the theory. Although one can regard the perspective in \cite{Kastor:2009wy} as being about variation of the parameter $\Lambda$, this perspective is not required, since $\Lambda$ can emerge, for example, as the constant of integration of a $4$-form gauge field (see, e.g., \cite{Creighton:1995au}).

To conclude, the cosmological constant in this context plays the role of a parameter in the action, rather than an integration constant characterizing a specific solution, and so its variation is going to change the theory in a similar way  the asymptotic value of the dilaton is doing when considering asymptotically flat black holes. Therefore, a natural interpretation could be that the same black hole is, in fact, considered in different environments with different cosmological constants and the energy is changing accordingly.

\begin{figure}[h]
	\centering
	\subfigure{\includegraphics[width=9.2 cm]{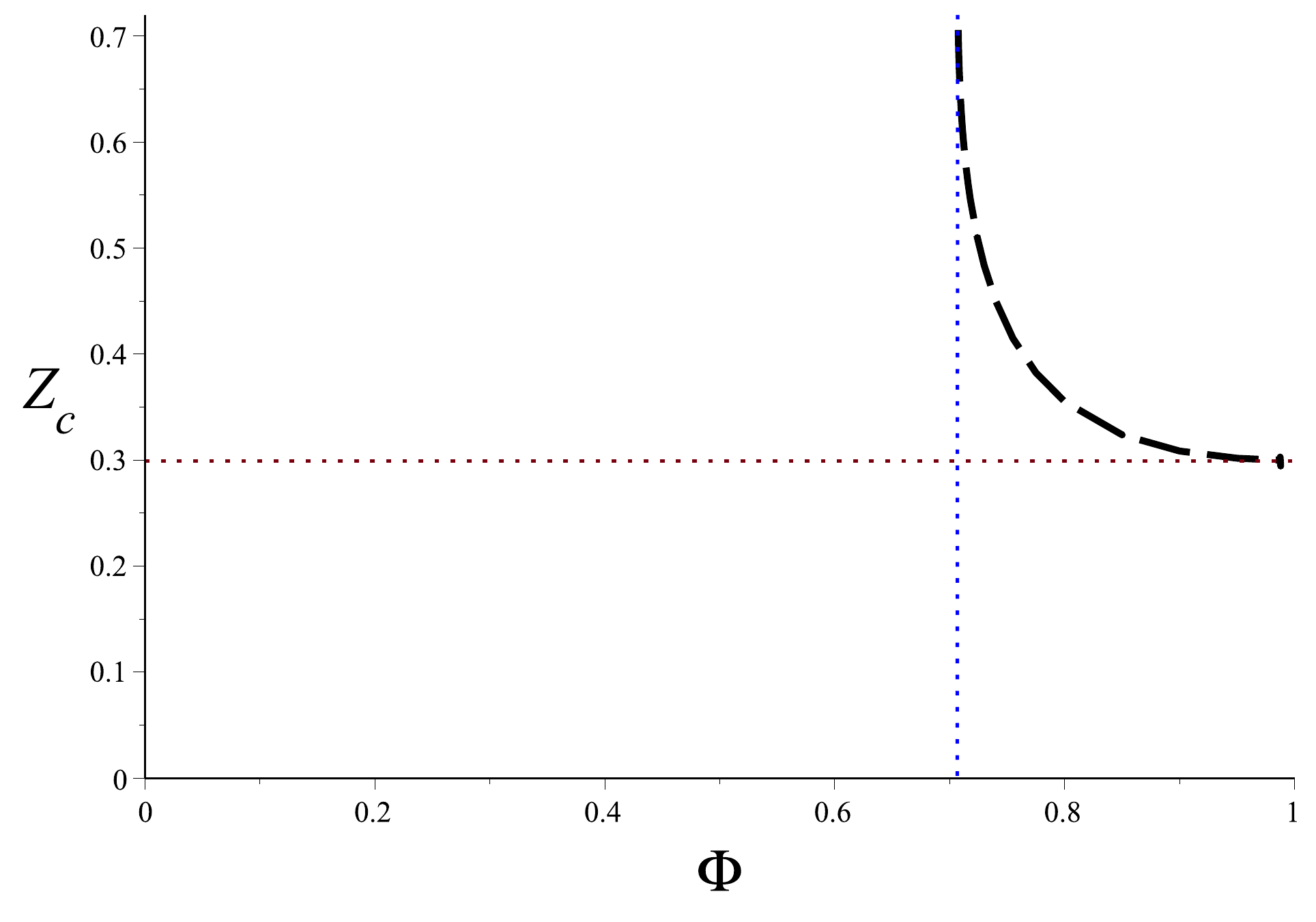}}
	\caption{\small Critical compressibility factor $Z_c\equiv \frac{P_cv_c}{T_c}$ vs the conjugate potential (dashed line), in the window for criticality, $\frac{1}{\sqrt{2}}<\Phi<1$. It is observed that $0.3\lesssim Z_c<\infty$. In the plot, the vertical dotted line is $\Phi=1/\sqrt{2}$, and the horizontal one to the limit $Z_c\approx 0.3$.}
\label{Zcrit}
\end{figure}

We have considered an exact regular hairy black hole solution in Einstein-Maxwell-dilaton theory with a non-trivial self-interaction for the scalar field. We have used the quasilocal formalism supplemented with counterterms to investigate the thermodynamics of this black hole. We were interested not only in the usual thermodynamics, but also in the extended thermodynamics when the cosmological constant becomes a dynamical parameter. Since we have presented in detail all the thermodynamic properties, we would like to only briefly discuss the main new characteristics due to the scalar field. Compared to RN-AdS black holes, the hairy black holes present criticality behaviour also in the grand canonical ensemble --- as we have already pointed out, there are some similarities with the canonical ensemble of RN-AdS black hole. An interesting difference, though, it is that the critical compressibility factor is not constant as in the case of RN-AdS black hole, but it ranges as $0.3\lesssim Z_c<\infty$, as shown in Figure \ref{Zcrit}.

\begin{figure}[h]
\centering
\subfigure{\includegraphics[width=7.2cm]
{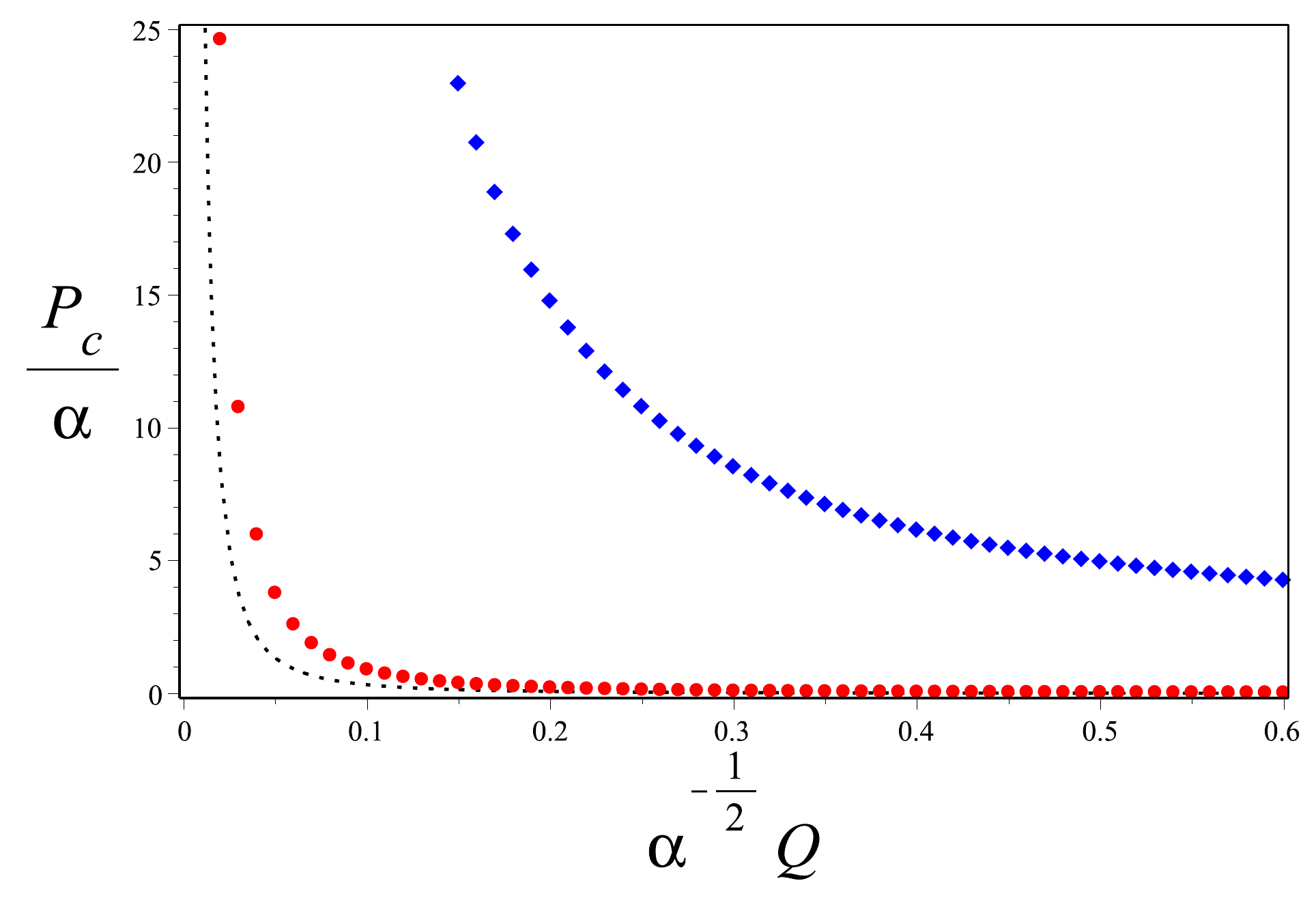}}	
\,\,\,\,
\subfigure{\includegraphics[width=7.1cm]
{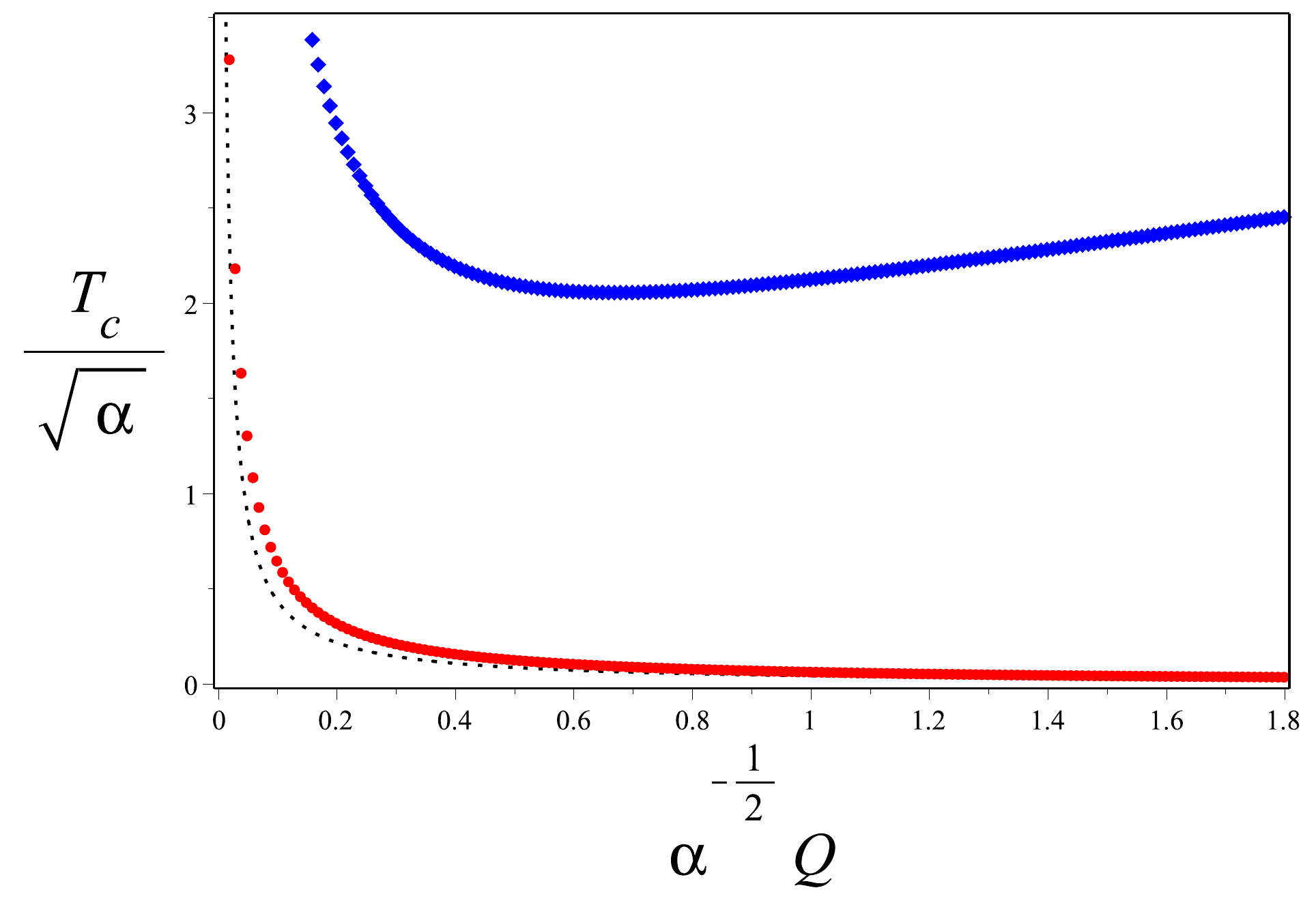}}
\caption{\small Critical pressure (Left) and the critical temperature (Right) as function of $\bar Q$. The red dots correspond to the RN-AdS-like critical points and the blue ones to the purely hairy ones. Black dotted lines correspond to the Reissner-N\"ordstrom-AdS black hole, given by equations (\ref{crRN}).
We see that  the RN-AdS-like critical points for hairy black holes closely match with the critical points for RN-AdS, merging with them as $Q/\sqrt{\alpha}$ gets large.}
\label{PcritTcrit}
\end{figure}

However, some important new features appear in the canonical ensemble when the charge is kept fixed. In this case the critical compressibility factor is not constant, but depends on the charge and so this should also be a characteristic of the dual field theory. Another important new characteristic is the appearance of a second critical point that does not have a counterpart in the RN-AdS case. In Figure \ref{PcritTcrit}, we plot independently the critical pressure and temperature as function of $\bar Q$ for both, the RN-AdS-like critical points (red dots) and the hairy critical points (blue ones).
One observation is that the RN-AdS-like critical points very closely fit  with the RN-AdS critical point described in the previous section (dotted curve, in Figure \ref{PcritTcrit}), and hence the name.

\section*{Acknowledgments}
The work of D.A. has been funded by the Fondecyt Regular Grant 1161418.  This work was supported in part by the  Natural Sciences and Engineering Research Council of Canada. RR was supported by the national Ph.D. scholarship Conicyt 21140024.


\end{document}